\documentclass[10pt, twocolumn]{IEEEtran}

\usepackage{amsmath}
\usepackage{amsfonts}
\usepackage{amssymb}
\usepackage{algorithm}
\usepackage{algorithmic}
\usepackage{graphicx}
\usepackage{verbatim}

\newtheorem{theorem}{Theorem}
\newtheorem{lemma}{Lemma}
\newtheorem{remark}{Remark}

\newtheorem{corollary}{Corollary}
\newtheorem{example}{Example}

\newcommand{\beq}{\begin{equation}}
\newcommand{\eeq}{\end{equation}}
\newcommand{\beqnn}{\begin{equation*}}
\newcommand{\eeqnn}{\end{equation*}}
\newcommand{\beqy}{\begin{eqnarray}}
\newcommand{\eeqy}{\end{eqnarray}}
\newcommand{\beqynn}{\begin{eqnarray*}}
\newcommand{\eeqynn}{\end{eqnarray*}}
\newcommand{\bit}{\begin{itemize}}
\newcommand{\eit}{\end{itemize}}
\newcommand{\ben}{\begin{enumerate}}
\newcommand{\een}{\end{enumerate}}
\newcommand{\bex}{\begin{example}}
\newcommand{\eex}{\end{example}}

\newcommand{\balg}[1]{\begin{algorithm} \caption{#1}}
\newcommand{\ealg}{\end{algorithm}}

\newcommand{\balgc}{\begin{algorithmic}[1]}
\newcommand{\ealgc}{\end{algorithmic}}

\newcommand{\bary}{\begin{array}}
\newcommand{\eary}{\end{array}}
\newcommand{\bmx}{\begin{bmatrix}}
\newcommand{\emx}{\end{bmatrix}}
\newcommand{\bsmx}{\left[\begin{smallmatrix}}
\newcommand{\esmx}{\end{smallmatrix}\right]}
\newcommand{\bmxc}[1]{\left[\begin{array}{@{}#1@{}}}
\newcommand{\emxc}{\end{array}\right]}
\newcommand{\bcn}{\begin{center}}
\newcommand{\ecn}{\end{center}}

\newcommand{\A}{\boldsymbol{A}}
\newcommand{\B}{\boldsymbol{B}}

\newcommand{\D}{\boldsymbol{D}}

\newcommand{\I}{\boldsymbol{I}}

\renewcommand{\P}{\boldsymbol{P}}

\renewcommand{\S}{\boldsymbol{S}}

\newcommand{\U}{\boldsymbol{U}}
\newcommand{\V}{\boldsymbol{V}}

\newcommand{\e}{\boldsymbol{e}}

\newcommand{\rr}{\boldsymbol{r}}

\renewcommand{\u}{\boldsymbol{u}}
\renewcommand{\v}{\boldsymbol{v}}
\newcommand{\w}{\boldsymbol{w}}
\newcommand{\x}{{\boldsymbol{x}}}
\newcommand{\y}{{\boldsymbol{y}}}

\newcommand{\0}{{\boldsymbol{0}}}

\newcommand{\1}{{\boldsymbol{1}}}

\newcommand{\bxi}{\boldsymbol{\xi}}

\usepackage[rgb]{xcolor}
\DeclareMathOperator*{\argmin}{arg\,min}
\DeclareMathOperator*{\argmax}{arg\,max}

\usepackage{slashbox}

\usepackage{psfrag}
\usepackage{amssymb}
\usepackage{amsmath}
\usepackage{cite}
\usepackage{graphics}
\usepackage{graphicx}
\usepackage{epsfig}
\usepackage{subfigure}
\usepackage{url}
\usepackage{amscd}
\usepackage{threeparttable}
\usepackage{enumerate}
\begin{document}
%
\title{A Sharp Condition for Exact Support Recovery With Orthogonal Matching Pursuit}

\author{Jinming~Wen, Zhengchun Zhou, Jian Wang, Xiaohu Tang, and  Qun Mo
\thanks{Copyright (c) 2015 IEEE. Personal use of this material is permitted. However, permission to use this material for any other purposes must be obtained from the IEEE by sending a request to pubs-permissions@ieee.org.}
\thanks{This work was supported by NSFC (No. 61672028, 11271010 and 11531013), and the Sichuan Provincial Youth Science and Technology Fund (No. 2015JQ0004 and 2016JQ0004), the fundamental research funds for the Central Universities, ``Programme Avenir
Lyon Saint-Etienne de l'Universit\'e de Lyon" in the framework of the programme
``Inverstissements d'Avenir" (ANR-11-IDEX-0007) and ANR through the HPAC project under
Grant ANR~11~BS02~013}

\thanks{This work was presented in part at the
IEEE International Symposium on Information Theory (ISIT 2016), Barcelona, Spain.}
\thanks{J.~Wen was with   ENS de Lyon, Laboratoire LIP (U. Lyon, CNRS, ENSL, INRIA, UCBL), Lyon 69007, France.
He is with  the Department of Electrical and Computer Engineering, University of Alberta, Edmonton T6G 2V4, Canada (e-mail: jinming1@ualberta.ca).}
\thanks{Z. Zhou (Corresponding Author) is with the School of Mathematics, Southwest Jiaotong University,
Chengdu 610031, China, (e-mail: zzc@home.swjtu.edu.cn). He is also with State Key Laboratory of Information Security
(Institute of Information Engineering, Chinese Academy
of Sciences, Beijing 100093).}
\thanks{J.~Wang is with the Department of Electrical and Computer Engineering, Duke University, USA, 27708,
(e-mail: wangjianeee@gmail.com).}
\thanks{X. Tang is with the Information Security and National Computing Grid
Laboratory, Southwest Jiaotong University, Chengdu 610031, China, (e-mail: xhutang@swjtu.edu.cn).}
\thanks{Q. Mo is with the Department of Mathematics, Zhejiang University,
Hangzhou 310027, China (e-mail: moqun@zju.edu.cn).}

%
}



\maketitle

\begin{abstract}
Support recovery of sparse signals from noisy measurements with orthogonal matching pursuit (OMP) has been extensively studied.
In this paper, we show that for any $K$-sparse signal $\x$,
if  a sensing matrix $\A$ satisfies the restricted isometry property (RIP) with restricted isometry constant (RIC)
$\delta_{K+1} < 1/\sqrt {K+1}$, then under some constraints on the minimum magnitude of nonzero elements of $\x$,
OMP exactly recovers the support of $\x$ from its measurements $\y=\A\x+\v$ in $K$ iterations,
where $\v$ is a noise vector that is $\ell_2$ or $\ell_{\infty}$ bounded.
This sufficient condition is sharp in terms of $\delta_{K+1}$ since
for any given positive integer $K$ and any $1/\sqrt{K+1}\leq \delta<1$,
there always exists a matrix $\A$ satisfying the RIP with $\delta_{K+1}=\delta$
for which OMP fails to recover a $K$-sparse signal $\x$ in $K$ iterations.
Also, our constraints on the minimum magnitude of nonzero elements of $\x$ are weaker than existing ones.
Moreover, we propose worst-case necessary conditions for the exact support recovery of $\x$,
characterized by the  minimum magnitude of the nonzero elements of $\x$.
\end{abstract}

\begin{IEEEkeywords}
Compressed sensing (CS), restricted isometry property (RIP), restricted isometry constant (RIC), orthogonal matching pursuit (OMP), support recovery.
\end{IEEEkeywords}

\section{Introduction}

\IEEEPARstart{I}{n} compressed sensing (CS), we frequently encounter the following linear model~\cite{CanT05,Don06,CohDD09,WenLZ15}:
\beq
\label{e:model}
\y=\A\x+\v,
\eeq
where $\x\in \mathbb{R}^n$ is an unknown $K$-sparse signal, 
(i.e., $|\text{supp}(\x)|\leq K$, where $\text{supp}(\x)=\{i:x_i\neq0\}$ is the support of $\x$
and $|\text{supp}(\x)|$ is the cardinality of $\text{supp}(\x)$.),
$\A\in \mathbb{R}^{m\times n}$ ($m\ll n$) is a known sensing matrix,
$\y\in \mathbb{R}^m$ contains the noisy observations (measurements),
and $\v \in \mathbb{R}^{m}$ is a noise vector.
There are several common types of noises, such as the $\ell_2$ bounded noise (i.e., $\|\v\|_2\leq \epsilon$ for some constant $\epsilon$
\cite{Fuc05,DonET06,Can08}), the $\ell_{\infty}$ bounded noise (i.e., $\|\A\v\|_{\infty}\leq \epsilon$ for some constant $\epsilon$~\cite{CaiW11}),
and the Gaussian noise (i.e., $v_i\sim \mathcal{N}(0, \sigma^2)$~\cite{CanT07}).
In this paper, we consider only the first two types of noises, as the analysis for these two types can be easily extended to the last one
by following some techniques in~\cite{CaiW11}.

One of the central goals of CS is to recover the sparse signal $\x$ on the basis of the sensing matrix $\A$ and the observations $\y$.
It has been demonstrated that under appropriate conditions on $\A$, the original signal $\x$ can be reliably recovered
via properly designed algorithms~\cite{CanRT06} \cite{Mol11}.
Orthogonal matching pursuit (OMP)~\cite{pati1993orthogonal} \cite{TroG07} is a widely-used greedy algorithm for performing the recovery task.
For any set $S\subset\{1,2,\cdots ,n\}$, let $\A_S$ denote the submatrix of $\A$ that contains only the columns indexed by $S$. Similarly, let $\x_S$ denote the subvector of $\x$ that  contains only the entries indexed by $S$. Then, the OMP algorithm  is formally described in Algorithm~\ref{a:OMP}.\footnote{If the maximum correlation in Step 2 occurs for multiple indices, break the tie randomly.}

\begin{algorithm}[t]
\caption{The OMP Algorithm~\cite{pati1993orthogonal}}  \label{a:OMP}
Input: $\y$, $\A$, and stopping rule.\\
Initialize: $k=0, \rr^0=\y, S_0=\emptyset$.\\
until the stopping rule is met
\begin{algorithmic}[1]
\STATE $k=k+1$,
\STATE $s^k=\argmax\limits_{1\leq i\leq n}|\langle \rr^{k-1},\A_i\rangle|$,
\STATE $S_k=S_{k-1}\bigcup\{s^k\}$,
\STATE $\hat{\x}_{S_k}=\argmin\limits_{x\in \mathbb{R}^{|S_k|}}\|\y-\A_{S_k}\x\|_2$,
\STATE $\rr^k=\y-\A_{S_k}\hat{\x}_{S_k}$.
\end{algorithmic}
Output: $\hat{\x}=\argmin\limits_{\x: \text{supp}(\x)=S_k}\|\y-\A\x\|_2$.
\end{algorithm}

A widely used framework for analyzing the recovery performance of the CS recovery algorithms is the restricted isometry property (RIP) \cite{CanT05}.
For an $m\times n$ matrix $\A$ and any integer $K$, the order-$K$ restricted isometry constant (RIC) $\delta_K$ is
defined as the smallest constant such that
\begin{equation}
\label{e:RIP}
(1-\delta_K)\|\x\|_2^2\leq \|\A\x\|_2^2\leq(1+\delta_K)\|\x\|_2^2
\end{equation}
for all $K$-sparse vectors $\x$.

For the noise-free case (i.e., when $\v = \0$), many RIC-based conditions have been proposed to guarantee the exact recovery of sparse signals via OMP.
It has respectively been shown in~\cite{DavW10} and~\cite{LiuT12} that $\delta_{K+1}<1/(3\sqrt{K})$ and
$\delta_{K+1}<1/\sqrt{2 K}$ are sufficient for OMP to recover any $K$-sparse signal $\x$ in $K$ iterations.
Later, the conditions have been improved to $\delta_{K+1}<1/(1+\sqrt{K})$~\cite{MoS12}~\cite{WanS12}
and further to $\delta_{K+1}<(\sqrt{4K+1}-1)/(2K)$~\cite{ChaW14}.
Recently, it has been shown that if $\delta_{K+1}<1/\sqrt{K+1}$,
OMP is guaranteed to exactly recover $K$-sparse signals $\x$ in $K$ iterations~\cite{Mo15}.
On the other hand, it has been conjectured in~\cite{DaiM09} that there exists a matrix $\A$ satisfying the RIP with $\delta_{K+1}\leq1/\sqrt{K}$
such that OMP fails to recover a $K$-sparse vector $\x$ in $K$ iterations.
This conjecture has been confirmed by examples provided in~\cite{MoS12}~\cite{WanS12}.
Furthermore, it has been reported in~\cite{WenZL13} \cite{Mo15} that for any given positive integer $K\geq 2$ and
any given $\delta$ satisfying $1/\sqrt{K+1}\leq \delta<1$, there always exist a $K$-sparse vector $\x$
and a matrix $\A$ satisfying the RIP with $\delta_{K+1}=\delta$
such that the OMP algorithm fails to recover $\x$ in $K$ iterations.
In other words, sufficient conditions for recovering $\x$ with $K$ steps of OMP cannot be weaker than $\delta_{K+1}<1/\sqrt{K+1}$, which
therefore implies that $\delta_{K+1}<1/\sqrt{K+1}$ is a sharp condition~\cite{Mo15}.

For the noisy case (i.e., when $\v\neq \0$), we are often interested in recovering the support of $\x$, i.e, $\text{supp}(\x)$.
Once $\text{supp}(\x)$ is exactly recovered, the underlying signal $\x$ can be easily estimated by ordinary least squares regression~\cite{CaiW11}.
It has been shown in \cite{SheL15} that under some constraint on the minimum magnitude of nonzero elements of $\x$ (i.e., $\min_{i\in\text{supp}(\x)}|x_i|$),
$\delta_{K+1} < 1/(\sqrt {K}+3 )$ is sufficient for OMP to  exactly recover $\text{supp}(\x)$ under both the $\ell_{2}$ and $\ell_{\infty}$ bounded noises.
The sufficient condition has been improved to $\delta_{K+1} < 1/(\sqrt {K}+1)$~\cite{WuHC13},
and the best existing condition in terms of $\delta_{K+1}$ is $\delta_{K+1}<(\sqrt{4K+1}-1)/(2K)$~\cite{ChaW14}.

In this paper, we investigate sufficient, and {\em worst-case} necessary conditions,
based on the RIC and $\min_{i\in\text{supp}(\x)}|x_i|$,
for recovering $\text{supp}(\x)$ with OMP under both $\ell_{2}$ and $\ell_{\infty}$ bounded noises.
Here, the worst-case necessity means that if it is violated, then there is (at least) one instance of $\{\A, \x, \v\}$
such that OMP fails to recover $\text{supp}(\x)$ from the noisy measurements $\y=\A\x+\v$~\cite{herzet2012exact}.
Specifically, our contributions can be summarized as follows.
\begin{enumerate}[i)]

\item
We show that if $\A$ and $\v$ in~\eqref{e:model} respectively satisfy the RIP with $\delta_{K+1} < 1/\sqrt {K+1}$
and $\|\v\|_2\leq \epsilon$,
then OMP with the stopping rule $\|\rr^k\|_2\leq \epsilon$
 exactly recovers $\text{supp}(\x)$ in $K$ iterations, provided that
$$
\min_{i\in\text{supp}(\x)}|x_i|>\frac{2\epsilon}{1-\sqrt {K+1 }\delta_{K+1}}.
$$
We also show that our constraint on $\min_{i\in\text{supp}(\x)}$ is weaker than existing ones.\footnote{These results were presented at the 2016 IEEE International Symposium on Information Theory (ISIT) conference~\cite{WenZWM16a}.} (Theorem \ref{t:l2}).

\item
We show that if $\A$ and $\v$ in~\eqref{e:model} respectively satisfy the RIP with
$\delta_{K+1} < 1/\sqrt {K+1}$ and $\|\A^T\v\|_{\infty}\leq \epsilon$, then OMP with the stopping rule
$$
\|\A^T\rr^k\|_{\infty}\leq \left(1+\sqrt{\frac{(1+\delta_{2})K}{1-\delta_{K+1}}}  \right)\epsilon
$$
 exactly recovers $\text{supp}(\x)$ in $K$ iterations, provided that
$$
\min_{i\in\text{supp}(\x)} \hspace{-.5mm} |x_i| \hspace{-.5mm} > \hspace{-.5mm} \frac{2}{1 \hspace{-.5mm} - \hspace{-.5mm} \sqrt {K \hspace{-.5mm} + \hspace{-.5mm} 1 }\delta_{K+1}} \hspace{-.5mm} \left(\hspace{-.5mm} 1\hspace{-.5mm} + \hspace{-.5mm} \sqrt{\frac{(1 \hspace{-.5mm} + \hspace{-.5mm} \delta_{2})K}{1-\delta_{K+1}}} \right)\epsilon.
$$
We also compare our constraint on $\min_{i\in\text{supp}(\x)}  |x_i|$ with existing results (Theorem \ref{t:linf}).

\item
We show that for any given positive integer $K$, $0< \delta<1/\sqrt{K+1}$, and $\epsilon>0$,
there always exist a sensing matrix $\A\in \mathbb{R}^{m\times n}$ satisfying the RIP with $\delta_{K+1}=\delta$,
a $K$-sparse vector $\x\in \mathbb{R}^{n}$ with
\[
\min_{i\in\text{supp}(\x)}|x_i|<\frac{\sqrt{1-\delta}\epsilon}{1-\sqrt {K+1 }\delta},
\]
and a noise vector $\v\in \mathbb{R}^{m}$ with $\|\v\|_2\leq \epsilon$,
such that OMP fails to recover $\text{supp}(\x)$ from $\y=\A\x+\v$ in $K$ iterations (Theorem \ref{t:l2nc}).

\item
We show that for any given positive integer $K$, $0<\delta<1/\sqrt{K+1}$ and $\epsilon>0$,
there always exist a sensing matrix $\A\in \mathbb{R}^{m\times n}$ satisfying the RIP with $\delta_{K+1}=\delta$,
a $K$-sparse vector $\x\in \mathbb{R}^{n}$ with
\[
\min_{i\in\text{supp}(\x)}|x_i|<\frac{2\epsilon}{1-\sqrt {K+1 }\delta},
\]
and a noise vector $\v\in \mathbb{R}^{m}$ with $\|\A^T\v\|_{\infty}\leq \epsilon$,
such that OMP fails to recover $\text{supp}(\x)$ from $\y=\A\x+\v$ in $K$ iterations (Theorem \ref{t:linfnc}).

\end{enumerate}

Since OMP may fail to recover a $K$-sparse signal with OMP in $K$ iterations
when $\A$ satisfies the RIP with $\delta\geq1/\sqrt{K+1}$ and $\v=\0$~\cite{WenZL13} \cite{Mo15},
sufficient conditions for recovering $\text{supp}(\x)$ with OMP in $K$ iterations in the noisy case cannot
be weaker than $\delta<1/\sqrt{K+1}$ (note that $\v=\0$ is the ideal case).
Hence,  our sufficient conditions summarized in i)--ii) are sharp {in terms of} the RIC.
Moreover, iii) and iv) indicate that
{for all $K$-sparse vectors $\x$ and sensing matrices $\A$ satisfying the RIP with $\delta_{K+1} < 1/\sqrt {K+1}$},
the worst-case necessary constraint on $\min_{i\in\text{supp}(\x)}|x_i|$ guaranteeing the exact recovery of $\text{supp}(\x)$
from \eqref{e:model} with $K$ iterations of OMP are
\[
\min_{i\in\text{supp}(\x)}|x_i|\geq\frac{\sqrt{1-\delta_{K+1}}\epsilon}{1-\sqrt {K+1 }\delta_{K+1}},
\]
and
\[
\min_{i\in\text{supp}(\x)}|x_i|\geq\frac{2\epsilon}{1-\sqrt {K+1 }\delta_{K+1}},
\]
under the $\ell_2$ and $\ell_{\infty}$ bounded noise, respectively.

The rest of the paper is organized as follows.
In Section~\ref{s:prelimiaries}, we introduce some notations that will be used throughout this paper. We also propose a
lemma which plays a central role in proving our new sufficient conditions.
In Section~\ref{s:main}, we present sufficient, and worst-case necessary conditions
for the exact support recovery of sparse signals with OMP under both the $\ell_{2}$ and $\ell_{\infty}$ bounded noises.
Finally, we conclude our paper in Section~\ref{s:con}.

\section{Preliminaries}
\label{s:prelimiaries}
\subsection{Notation}
We first define some notations that will be used throughout this paper.
Let $\mathbb{R}$ be the real field. Boldface lowercase letters denote column vectors, and boldface uppercase letters denote matrices.
For a vector $\x$, $\x_{i:j}$ denotes the subvector of $\x$ formed by entries $i, i+1, \cdots,j$.
Let $\e_k$ denote the $k$-th column of the identity matrix $\I$
and $\0$ denote a zero matrix or a zero column vector.
Denote $\Omega=\text{supp}(\x)$ and $|\Omega|$ be the cardinality of $\Omega$,
then for any $K$-sparse signal $\x$, $|\Omega|\leq K$.
For any set $S$, denote $\Omega \setminus S=\{i|i\in\Omega,i \notin S\}$.
Let $\Omega^c$ and $S^c$ denote the complement of $\Omega$ and $S$, respectively,
i.e., $\Omega^c=\{1,2,\ldots ,n\}\setminus \Omega$, and $S^c=\{1,2,\ldots ,n\}\setminus S$.
Let $\A_S$ denote the submatrix of $\A$ that  contains only the columns indexed by $\S$.
Similarly, let $\x_S$ denote the subvector of $\x$ that  contains only the entries indexed by $\S$.
For any matrix $\A_S$ of full column-rank, let $\P_S=\A_S(\A_S^T\A_S)^{-1}\A_S^T$
and $\P^{\bot}_S=\I-\P_S$ denote the projector and orthogonal complement projector onto the column space of $\A_S$,  respectively, where $\A_S^T$ stands for the transpose of $\A_S$.

\subsection{A useful lemma}\label{sec:lem}

We present the following lemma, which is one of the central results of this paper and
will play a key role in proving our sufficient conditions for support recovery with OMP.
\begin{lemma}
\label{l:main}
Suppose that $\A$ in~\eqref{e:model} satisfies the RIP of order $K+1$ and $S$ is a proper subset of $\Omega$
(i.e., $S \subset \Omega$ with $|S|<|\Omega|$). Then,
\begin{eqnarray}\label{e:main}
\lefteqn{\|\A_{\Omega\setminus S}^T\P^{\bot}_S\A_{\Omega\setminus S}\x_{\Omega\setminus S}\|_{\infty}
-\|\A_{\Omega^c}^T\P^{\bot}_S\A_{\Omega\setminus S}\x_{\Omega\setminus S}\|_{\infty}} ~~~~~~~~~\nonumber\\
&& \geq \frac{(1-\sqrt {|\Omega|-|S|+1 }\delta_{|\Omega|+1})\|\x_{\Omega\setminus S}\|_2}{\sqrt{|\Omega|-|S|}}.
\end{eqnarray}
\end{lemma}
{\em Proof.}
See Appendix~\ref{ss:lemma1}. ~~~~~~~~~~~~~~~~~~~~~~~~~~~~~~~~~~~~~~~~~
$\Box$

Note that in the noise-free case (i.e., when $\v=\0$), Lemma~\ref{l:main} can be directly connected to the selection rule of OMP.
Specifically, if we assume that $S = S_k \subset \Omega$ for some $0 \leq k < |\Omega|$ (see Algorithm~\ref{a:OMP} for the definition of $S_k$), then
\begin{equation}
\label{e:rknoiseless}
\P^{\bot}_S\A_{\Omega\setminus S}\x_{\Omega\setminus S} \hspace{-.5mm} = \hspace{-.5mm} \P^{\bot}_{S_k} \A_{\Omega \setminus S_k}\x_{\Omega\setminus S_k} \hspace{-.5mm} = \hspace{-.5mm} \P^{\bot}_{S_k} \A \x \hspace{-.5mm} = \hspace{-.5mm} \P^{\bot}_{S_k} \y=\rr^k
\end{equation}
and \eqref{e:main} can be rewritten as
\begin{eqnarray}
\label{e:mainnoiseless}
\lefteqn{\max_{j\in\Omega\setminus S_k}|\langle\A_j,\rr^k\rangle|
-\max_{j\in\Omega^c}|\langle\A_j,\rr^k\rangle|}~~~ \nonumber\\
&& \geq \frac{(1-\sqrt {|\Omega|-k+1 }\delta_{|\Omega|+1})\|\x_{\Omega\setminus S_k}\|_2}{\sqrt{|\Omega|-k}}.
\end{eqnarray}
Clearly, \eqref{e:mainnoiseless} characterizes a lower bound on the difference between
the maximum value of the OMP decision-metric for the columns belonging to $\Omega$ and that for the columns belonging to $\Omega^c$.
Since $\|\x_{\Omega\setminus S_k}\|_2>0$ for $k<|\Omega|$, \eqref{e:mainnoiseless} implies that OMP chooses
a correct index among $\Omega$ in the ($k + 1$)-th iteration as long as $\delta<1/\sqrt{|\Omega| - k +1}$.
Thus, by induction, one can show that OMP exactly recovers $\Omega$ in $K$ iterations under $\delta<1/\sqrt{K+1}$,
which matches the result in \cite{Mo15}.

In the noisy case (i.e., when $\v\neq\0$), by assuming that $S = S_k \subset \Omega$ for some $0 \leq k < |\Omega|$, we have
\begin{equation}
\label{e:rknoisy}
\P^{\bot}_S\A_{\Omega\setminus S}\x_{\Omega\setminus S}   \hspace{-.5mm} = \hspace{-.5mm} \P^{\bot}_{S_k} \A \x = \P^{\bot}_{S_k} \y - \P^{\bot}_{S_k} \v = \rr^k - \P^{\bot}_{S_k} \v.
\end{equation}
Due to the extra term $\P^{\bot}_{S_k} \v$ in \eqref{e:rknoisy}, however, we cannot directly obtain~\eqref{e:mainnoiseless} from \eqref{e:main}.
Nevertheless, by applying \eqref{e:rknoisy} (i.e., the relationship between $\P^{\bot}_S\A_{\Omega\setminus S}\x_{\Omega\setminus S}$ and $\rr^k$),
one can implicitly obtain from \eqref{e:main} a lower bound for
\[
\max_{j\in\Omega\setminus S_k}|\langle\A_j,\rr^k\rangle|
-\max_{j\in\Omega^c}|\langle\A_j,\rr^k\rangle|
\]
by utilizing $\|\x_{\Omega\setminus S_k}\|_2$, $\delta_{|\Omega|+1}$ and $\v$.
The lower bound, together with some constraints on $\min_{i\in \Omega}  |x_i|$,  allows to build sufficient conditions for OMP
to select an index belonging to $\Omega$ at the ($k + 1$)-th iteration.
See more details in Section III-A.

\begin{remark}
Lemma \ref{l:main} is motivated by~\cite[Lemma II.2]{Mo15}.
But these two lemmas have a key distinction. Specifically, while \cite[Lemmas II.2]{Mo15} showed that
\[
\|\A_{\Omega}^T\A\x\|_{\infty}>\|\A_{\Omega^c}^T\A\x\|_{\infty},
\]
Lemma~\ref{l:main} quantitatively characterizes a lower bound on
\[
\|\A_{\Omega\setminus S}^T\P^{\bot}_S\A_{\Omega\setminus S}\x_{\Omega\setminus S}\|_{\infty}
-\|\A_{\Omega^c}^T\P^{\bot}_S\A_{\Omega\setminus S}\x_{\Omega\setminus S}\|_{\infty},
\]
which is stronger in the following aspects:
\begin{enumerate}[i)]

\item Lemma~\ref{l:main} is more general and embraces \cite[Lemma II.2]{Mo15} as a special case. To be specific, Lemma \ref{l:main} holds for any $S$ which is a proper subset of $\Omega$,
while \cite[Lemma II.2]{Mo15} only works for the case where $S=\emptyset$.
In fact, the generality of Lemma~\ref{l:main} (i.e., it works for any $S\subset \Omega$) is of vital importance for the noisy case analysis of OMP.
Indeed, due to the noise involved, the recovery condition for the first iteration of OMP does not apply to the succeeding iterations.\footnote{This is in contrast to the noise-free case where the condition for the first iteration of OMP immediately applies to
 the succeeding iterations because the residual of those iterations can be viewed as the {\em modified measurements} of $K$-sparse signals
 with the same sensing matrix $\A$; See, e.g., \cite[Lemma 6]{WanS12}.}
 Thus, we need to consider the recovery condition for every individual iteration of OMP, which, as will be seen later, essentially corresponds to the cases of $S = S_k$, $k = 0, 1, \cdots, K - 1$, in Lemma~\ref{l:main}.

\item
In contrast to \cite[Lemma II.2]{Mo15} that is applicable to the noise-free case only, the lower bound in Lemma~\ref{l:main} works for both the noise-free as well as the noisy case (as indicated above).
Specifically, as will be seen in Appendix~B, by applying this quantitative lower bound with $S = S_k$, together with the relationship between the residual and $\P^{\bot}_{S_k} \A_{\Omega \setminus S_k}\x_{\Omega\setminus S_k}$ (see \eqref{e:rknoisy}), we are able to get a precise characterization of the difference between
the maximum value of OMP decision metrics for correct and incorrect columns in the noisy case, from which the sufficient condition guaranteeing a correct selection immediately follows.

\item Compared to \cite[Lemma II.2]{Mo15}, Lemma \ref{l:main} gives a sharper lower bound on the difference between
the maximum value of the OMP decision-metric for the columns belonging to $\Omega$ and that for the columns belonging to $\Omega^c$.
Specifically, \cite[Lemma II.2]{Mo15} showed that $\|\A_{\Omega}^T\A\x\|_{\infty}-\|\A_{\Omega^c}^T\A\x\|_{\infty}$ is lower bounded by zero
when $\A$ satisfies the RIP with $\delta_{K+1}<1/\sqrt{K+1}$.
Whereas, applying $S = \emptyset$ in \eqref{e:main} yields
\begin{equation}
\|\A_{\Omega}^T \A \x \|_{\infty} \hspace{-.5mm}
- \hspace{-.25mm} \|\A_{\Omega^c}^T \A \x \|_{\infty}\nonumber\\
\hspace{-.5mm}  \geq  \hspace{-.5mm}  (1-\sqrt {|\Omega|\hspace{-.5mm}  + \hspace{-.5mm} 1 }\delta_{|\Omega|+1})\hspace{-.25mm}   \min_{i\in\Omega} \hspace{-.25mm} |\x_{i}|,
\end{equation}
where the right-hand side can be much larger than zero under the same RIP assumption.

\end{enumerate}
\end{remark}

\section{Main Analysis}
\label{s:main}

In this section, we will show that if a sensing matrix $\A$
satisfies the RIP with $\delta_{K+1}<1/\sqrt{K+1}$, then under some constraints on
$\min_{i\in\Omega}|x_i|$, OMP exactly recovers $\text{supp}(\x)$ in
$K$ iterations from the noisy measurements $\y=\A\x+\v$.
We will also present worst-case necessary conditions on $\min_{i\in\Omega}|x_i|$ for the exact recovery of $\text{supp}(\x)$.

\subsection{Sufficient condition}
\label{ss:suff}
We consider both $\ell_{2}$ and $\ell_{\infty}$ bounded noises. The
following theorem gives a sufficient condition for the exact support
recovery with OMP under the $\ell_2$ bounded noise.

\begin{theorem}
\label{t:l2}
Suppose that $\A$ and $\v$ in~\eqref{e:model} respectively satisfy the RIP with
\beq
\label{e:delta}
\delta_{K+1} < \frac{1}{\sqrt {K+1 }},
\eeq
and $\|\v\|_2\leq \epsilon$.
Then OMP with the stopping rule $\|\rr^k\|_2\leq \epsilon$ exactly recovers the support $\Omega$ of any $K$-sparse signal $\x$ from $\y=\A\x+\v$ in $|\Omega|$ iterations, provided that
\beq
\label{e:l2}
\min_{i\in\Omega}|x_i|>\frac{2\epsilon}{1-\sqrt {K+1 }\delta_{K+1}}.
\eeq
\end{theorem}
{\em Proof.}
See Appendix~\ref{ss:l2}. ~~~~~~~~~~~~~~~~~~~~~~~~~~~~~~~~~~~~~~~~~
$\Box$

If $\v=\0$, then we can set $\epsilon=0$ which implies that \eqref{e:l2} holds.
Thus we have the following result, which coincides with~\cite[Theorem III.1]{Mo15}.
\begin{corollary}
\label{c:l2}
If $\A$ and $\v$ in~\eqref{e:model} respectively satisfy the RIP with~\eqref{e:delta} and $\v = 0$,
then OMP exactly recovers all $K$-sparse signals $\x$ from $\y = \A \x$ in $K$ iterations.
\end{corollary}

\begin{remark}
In~\cite{WenZL13} and \cite{Mo15}, it has been shown that for any given
integer $K\geq 2$ and for any $1/\sqrt{K+1}\leq \delta<1$, there always
exist a $K$-sparse vector $\x$ and a sensing matrix $\A$ with
$\delta_{K+1}=\delta$, such that the OMP algorithm fails to recover $\x$ from \eqref{e:model}
(note that this statement also holds for $K=1$). Therefore, the
sufficient condition in Theorem~\ref{t:l2} for the exact support recovery with OMP is sharp in terms of $\delta_{K+1}$.
\end{remark}

It might be interesting to compare our condition with existing results.
In~\cite{SheL15,WuHC13,ChaW14}, similar recovery conditions have
been proposed for the OMP algorithm under the assumption that the
sensing matrices $\A$ have normalized columns (i.e., $\|\A_i\|_2=1$
for $i=1,2,\ldots, n$). Comparing with these conditions, our condition given by
Theorem~\ref{t:l2} is more general as it works for sensing
matrices whose column $\ell_2$-norms are not necessarily equal to 1. More importantly, our result
is less restrictive than those in~\cite{SheL15,WuHC13,ChaW14} with respect to both $\delta_{K+1}$ and $\min_{i\in\Omega}|x_i|$. To
illustrate this, we compare our condition with that
in~\cite{ChaW14}, which is the best result to date.
In~\cite{ChaW14}, it was shown that if the sensing matrix $\A$ is
column normalized and satisfies the RIP with
$$
\delta_{K+1}<\frac{\sqrt{4K+1}-1}{2K}
$$
and the noise vector $\v$ satisfies $\|\v\|_2\leq \epsilon$, then
the OMP algorithm with the stopping rule $\|\rr^k\|_2\leq \epsilon$ exactly recovers
the support $\Omega$ of any $K$-sparse signal $\x$ from $\y = \A \x
+ \v$ in $K$ iterations, provided that
$$
\min_{i\in\Omega}|x_i|>\frac{(\sqrt{1+\delta_{K+1}}+1)\epsilon}{1-\delta_{K+1}-\sqrt{1-\delta_{K+1}}\sqrt{K}\delta_{K+1}}.
$$

To show our condition in Theorem~\ref{t:l2} is less restrictive, it suffices
to show that \beq \label{e:compl2ric}
\frac{\sqrt{4K+1}-1}{2K}<\frac{1}{\sqrt{K+1}} \eeq
and that
\begin{eqnarray}
  \label{e:compl2x}
\frac{(\sqrt{1+\delta_{K+1}}+1)\epsilon}{1-\delta_{K+1}-\sqrt{1-\delta_{K+1}}\sqrt{K}\delta_{K+1}}
>\frac{2\epsilon}{1-\sqrt {K+1 }\delta_{K+1}}.\hspace{-2mm}
\end{eqnarray}
To show~\eqref{e:compl2ric}, we need to show
$$
\sqrt{(4K+1)(K+1)}<2K+\sqrt {K+1 },
$$
which is equivalent to
$$
4K^2+5K+1<4K^2+K+1+4K\sqrt {K+1 }.
$$
Since $K\geq1$, the aforementioned equation holds, so~\eqref{e:compl2ric} holds.

We next fucus on the proof of~\eqref{e:compl2x}. Since $
\sqrt{1+\delta_{K+1}}+1 > 2,$ it is clear that \eqref{e:compl2x}
holds if
$$
1-\delta_{K+1}-\sqrt{1-\delta_{K+1}}\sqrt{K}\delta_{K+1}<1-\sqrt {K+1 }\delta_{K+1}.
$$
Equivalently,
\beq \label{e:compl2x1}
1+\sqrt{1-\delta_{K+1}}\sqrt{K}>\sqrt{K+1}.
\eeq
Obviously,~\eqref{e:compl2x1} holds if
$$
\sqrt{1-\delta_{K+1}}>\frac{\sqrt{K+1}-1}{\sqrt{K}},
$$
which is equivalent to
$$
\delta_{K+1}<\frac{2(\sqrt{K+1}-1)}{K}.
$$
By~\eqref{e:delta}, it suffices to show
$$
\frac{1}{\sqrt{K+1}}<\frac{2(\sqrt{K+1}-1)}{K}.
$$
By some simple calculations, one can show that the aforementioned inequality holds.
Thus, \eqref{e:compl2x} holds under~\eqref{e:delta}.

Now we turn to the case where the noise vector $\v$ is
$\ell_{\infty}$ bounded.

\begin{theorem}
\label{t:linf} Suppose that $\A$ and $\v$
in~\eqref{e:model} respectively satisfy the RIP with~\eqref{e:delta} and
$\|\A^T\v\|_{\infty}\leq \epsilon$. Then OMP with the stopping rule
 \beq
 \label{e:stopcinf}
\|\A^T \rr^k\|_{\infty}\leq
\left(1+\sqrt{\frac{1+\delta_{2}}{1-\delta_{K+1}}}\sqrt{K}\right)\epsilon
\eeq
exactly recovers the support $\Omega$ of any $K$-sparse signal
$\x$ from $\y=\A\x+\v$ in $|\Omega|$ iterations, provided that\footnote{If the columns of $\A$ exhibit a unit $\ell_2$ norm, then \eqref{e:stopcinf} and \eqref{e:linf}
can be respectively relaxed to
$
\|\A^T\rr^k\|_{\infty}\leq \left(1+\frac{\sqrt{K}}{\sqrt{1-\delta_{K+1}}}\right)\epsilon
$
and
$
\min\limits_{i\in\Omega}|x_i|>\frac{2}{1-\sqrt {K+1 }\delta_{K+1}}\left(1+\frac{\sqrt{K}}{\sqrt{1-\delta_{K+1}}}\right)\epsilon.
$}
\begin{eqnarray}
  \label{e:linf}
\min_{i\in\Omega}|x_i| \hspace{-.25mm}>\hspace{-.25mm}\frac{2}{1\hspace{-.25mm}-\hspace{-.25mm}\sqrt {K\hspace{-.25mm}+\hspace{-.25mm}1
}\delta_{K+1}} \hspace{-.25mm}
\left(\hspace{-.25mm}1\hspace{-.25mm}+\hspace{-.25mm}\sqrt{\frac{1+\delta_{2}}{1\hspace{-.25mm}-\hspace{-.25mm}\delta_{K+1}}}\sqrt{K}
\right)\hspace{-.25mm}\epsilon. \hspace{-2mm}
\end{eqnarray}

\end{theorem}

{\em Proof.}
See Appendix~\ref{ss:linf}. ~~~~~~~~~~~~~~~~~~~~~~~~~~~~~~~~~~~~~~~~~
$\Box$

\begin{remark}
While in~\cite{WuHC13,ChaW14,SheL15}, \beq \label{e:stopcinf3}
\|\A^T\rr^k\|_{\infty}\leq \epsilon \eeq was used as the stopping
rule of OMP, we would like to note that \eqref{e:stopcinf}
in Theorem~\ref{t:linf}
cannot be replaced by~\eqref{e:stopcinf3}. Otherwise, OMP may choose
indices that do not belong to $\Omega$, no matter how large $\min_{i\in\Omega}|x_i|$ is.
To illustrate this, we give an example, where for simplicity
we consider sensing matrix $\A$ with $\ell_2$-normalized columns.

\begin{example}
\label{ex:ex1} For any $0<\delta<1$ and any
$a>\frac{1+\delta}{1-\delta}$, let
\beqnn
\hspace{-.1mm}
\A= \bmx
\sqrt{1-\delta^2}&0\\
\delta&1
\emx,\,
\x=
\bmx
a\\
0
\emx,
\v=
\bmx
-\frac{2\delta}{\sqrt{1-\delta^2}}\\
1 \emx,\, \epsilon=1. \hspace{-2mm}
\eeqnn
Then, $\x$ is an $1$-sparse
vector supported on $\Omega=\{1\}$ and $$\min_{i\in\Omega}|x_i| =
a.$$ $\A$ has unit $\ell_2$-norm columns. It is easy to verify that
the singular values of $\A^T\A$ are $1\pm\delta$, which, by
the definition of the RIC, implies that $\delta_{2}=\delta$. Moreover,
since
$$
\A^T\v=
\bmx
-\delta\\
1 \emx,
$$
we have $ \|\A^T\v\|_{\infty}\leq\epsilon.$ In the following, we
show that  if~\eqref{e:stopcinf3} is used as the stopping rule,
then OMP finally returns the index set $\{1,2\}$, no matter how large
$a$ is.

By~\eqref{e:model}, we have
$$
\y=
\bmx
a\sqrt{1-\delta^2}-\frac{2\delta}{\sqrt{1-\delta^2}}\\
a\delta+1
\emx.
$$
Thus,
$$
\A_1^T\y=a-\delta, \,\A_2^T\y=a\delta+1.
$$
Since $a>\frac{1+\delta}{1-\delta}$, we have $\A_1^T\y>\A_2^T\y$.
Thus by the selection rule of OMP (see Algorithm~\ref{a:OMP}),
$S_1=\{1\}$ is identified and consequently,
$$
\hat{x}_{S_1}=(\A_1^T\A_1)^{-1}\A_1^T\y=a-\delta
$$
and the residual is updated as
$$
\rr^1=\y-\A_1\hat{x}_{S_1}=
\bmx
\delta\sqrt{1-\delta^2}-\frac{2\delta}{\sqrt{1-\delta^2}}\\
1+\delta^2
\emx.
$$

By some calculations, we obtain
$$
\A^T\rr^1=
\bmx
0\\
1+\delta^2 \emx,
$$
which implies that $\|\A^T\rr^1\|_{\infty}=1+\delta^2>\epsilon$ so
that the stopping condition~\eqref{e:stopcinf3} does not satisfy. Hence, the
OMP algorithm will continue to the second iteration and will
eventually return the index set $\{1,2\}$.
\end{example}
\end{remark}

Again, by \cite{WenZL13} and \cite{Mo15}, the sufficient condition given in Theorem~\ref{t:linf} is sharp in terms of $\delta_{K+1}$.
We mention that similar constraints on $\min_{i\in\Omega}|x_i|$ have
been proposed in~\cite{WuHC13,ChaW14,SheL15}. However, since those
results were based on a different stopping rule
(i.e.,~\eqref{e:stopcinf3}), we do not give a comparison of our
constraint to those results.

\subsection{Worst-case necessary condition} \label{sec:nes}

In the above subsection, we have presented sufficient conditions
guaranteeing exact support recovery of sparse signals with OMP.
In this subsection, we investigate worst-case necessary conditions for the exact
support recovery. Like the sufficient conditions, our necessary
conditions are given in terms of the RIC as well as the minimum
magnitude of the nonzero elements of input signals.
Note that OMP may fail to recover a $K$-sparse signal $\x$ from $\y
= \A \x + \v$ if $\delta_{K+1} \geq 1/\sqrt{K+1}$, even
in the noise-free case \cite{WenZL13} \cite{Mo15}. Hence, $\delta_{K+1} < 1/\sqrt{K+1}$
naturally becomes a necessity for the noisy case.
Therefore,  in deriving the worst-case necessary condition on $\min_{i\in\Omega}|x_i|$,
we consider only the matrices $\A$ satisfying the RIP with $\delta_{K+1}<1/\sqrt{K+1}$.

We first restrict our attention to the case of $\ell_2$ bounded noises.

\begin{theorem}
\label{t:l2nc}
For any given $\epsilon>0$, positive integer $K$, and 
\beq
\label{e:deltanc}
0< \delta<\frac{1}{\sqrt{K+1}},
\eeq
there
always exist a matrix $\A\in \mathbb{R}^{m\times n}$ satisfying the RIP with $\delta_{K+1}=\delta$, a
$K$-sparse vector $\x\in \mathbb{R}^{n}$ with
\[
\min_{i\in\Omega}|x_i|<\frac{\sqrt{1-\delta}\epsilon}{1-\sqrt {K+1 }\delta},
\]
and a noise vector $\v\in \mathbb{R}^{m}$  with $\|\v\|_2\leq \epsilon$, such that OMP
fails to recover $\Omega$ from $\y = \A \x + \v$ in $K$
iterations.
\end{theorem}

{\em Proof.}
See Appendix~\ref{ss:l2nc}. ~~~~~~~~~~~~~~~~~~~~~~~~~~~~~~~~~~~~~~~~~
$\Box$

\begin{remark}
\label{remark:l2} One can immediately obtain from Theorem \ref{t:l2nc} that under the $\ell_2$ bounded noise,
a worst-case necessary condition (recall that the worst-case necessity means that if it is violated,
then there is (at least) one instance of $\{\A, \x, \v\}$
such that OMP fails to recover $\text{supp}(\x)$ from the noisy measurements $\y=\A\x+\v$~\cite{herzet2012exact}.)
on $\min_{i\in\Omega}|x_i|$ for OMP is:
\begin{eqnarray}
\label{e:l2nece}
\min_{i\in\Omega}|x_i|\geq\frac{\sqrt{1-\delta_{K+1}}\epsilon}{1-\sqrt {K+1 }\delta_{K+1}}.
\end{eqnarray}
Here, we would like to note that the {\it worst-case} necessity does not mean that  for $\A$, $\x$ and $\v$,
\eqref{e:l2nece} has to be satisfied to ensure the exact support recovery.
In fact, it can be shown that OMP may
be able to recover $\text{supp}(\x)$ in $K$ iterations when~\eqref{e:l2nece} does not hold. One such example is given as follows.
\begin{example}
\label{ex:ex2}
For any given $0<\delta<1/\sqrt{2}$ and
\[
0<a<\frac{\sqrt{2(1-\delta)}}{1-\sqrt {K+1 }\delta},
\]
let
\[
\A=\delta\I_2,\quad
\x=
\bmx
a\\
0
\emx,\quad
\v=
\bmx
1\\
1
\emx,\quad \epsilon=\sqrt{2}.
\]
Then, $\A$ satisfies the RIP with $\delta_{2}=\delta<1/\sqrt{2}$. Moreover, $\x$ is $1$-sparse
and does not satisfy \eqref{e:l2nece}. However,
one can check that OMP with the stopping condition $\|\rr^k\|_2\leq
\epsilon$ exactly recovers $\Omega$ in just one iteration.
\end{example}
\end{remark}

\begin{remark}
We mention that the worst-case necessary condition for the exact support recovery with OMP has also been studied in~\cite{Wan15},
in which the author characterized the worst-case necessity using the signal-to-noise ratio (SNR).
However, their result concerned only the sensing matrices $\A$ whose singular values are ones.
In comparison, our condition is more general and works for all sensing matrices $\A$ satisfying the RIP with $\delta_{K+1}<1/\sqrt{K+1}$.
\end{remark}

Next, we proceed to the worst-case necessity analysis for the case where the noise is
$\ell_{\infty}$ bounded.

\begin{theorem}
\label{t:linfnc}
For any given $\epsilon>0$, positive integer $K$, and $\delta$ satisfying
\eqref{e:deltanc}, there always exist a matrix $\A\in
\mathbb{R}^{m\times n}$ satisfying the RIP with $\delta_{K+1}=\delta$, a
$K$-sparse vector $\x\in \mathbb{R}^{n}$ with
\[
\min_{i\in\Omega}|x_i|<\frac{2\epsilon}{1-\sqrt {K+1 }\delta},
\]
and a noise vector $\v\in \mathbb{R}^{m}$ with
$\|\A^T\v\|_{\infty}\leq \epsilon$, such that OMP  fails to recover
$\Omega$ from $\y = \A \x + \v$ in $K$ iterations.
\end{theorem}
{\em Proof.}
See Appendix~\ref{ss:linfnc}. ~~~~~~~~~~~~~~~~~~~~~~~~~~~~~~~~~~~~~~~~~
$\Box$

\begin{remark}
Similar to the case of the $\ell_{2}$ bounded noise, Theorem~\ref{t:linfnc}
implies a worst-case necessary condition on $\min_{i\in\Omega}|x_i|$, for exactly recovering
$\text{supp}(\x)$ with OMP under the $\ell_{\infty}$ bounded noise is:
\begin{eqnarray}
\label{e:linfnece}
\min_{i\in\Omega}|x_i|\geq\frac{2\epsilon}{1-\sqrt {K+1 }\delta_{K+1}}.
\end{eqnarray}
Again, we note that \eqref{e:linfnece} applies to the worst case. For general cases, however, OMP may be able to
exactly recover $\Omega$ without this requirement. See a toy example as follows.
\begin{example}
\label{ex:ex3}
For any given $0<\delta<1/\sqrt{2}$ and
\[
0<a<\frac{2\sqrt{2}\delta.}{1-\sqrt {K+1 }\delta},
\]
let
\[
\A=\delta\I_2,\quad
\x=
\bmx
a\\
0
\emx,\quad
\v=
\bmx
1\\
1
\emx,\quad~\epsilon=\sqrt{2}\delta.
\]
Then, $\A$ satisfies the RIP with $\delta_{2}=\delta<1/\sqrt{2}$. $\x$ is
$1$-sparse and does not satisfy \eqref{e:linfnece}. Furthermore, one
can easily show that the OMP algorithm can exactly recover
$\Omega$ in one iteration when the stopping rule in
\eqref{e:stopcinf} is used.
\end{example}
\end{remark}

Finally, we would like to mention that while our sufficient conditions are sharp in terms of the RIC,
there is still some gap between the sufficient and the worst-case necessary constraints on $\min_{i\in\text{supp}(\x)}|x_i|$.
In particular, for the  $\ell_2$ bounded noise, the gap between conditions~\eqref{e:l2} and \eqref{e:l2nece} is relatively small,
demonstrating the tightness of the sufficient condition~\eqref{e:l2}.
For the $\ell_\infty$ bounded noise, however, the gap between conditions \eqref{e:linf} and~\eqref{e:linfnece} can be large
since the expression $\left(1 + \sqrt{\frac{1+\delta_{2}}{1 - \delta_{K+1}}}\sqrt{K}\right)$ on the right-hand side of~\eqref{e:linf}
can be much larger than one for a support cardinality $K$ that is large enough.
Whether it is possible to bridge this gap is an interesting open question.

\section{Conclusion}
\label{s:con} In this paper, we have studied sufficient conditions
for the exact support recovery of sparse signals from noisy
measurements by OMP. For both $\ell_2$ and $\ell_{\infty}$ bounded
noises, we have shown that if the RIC of a sensing matrix $\A$
satisfies the RIP with $\delta_{K+1} < 1/\sqrt {K+1}$, then under some conditions
on the minimum magnitude of nonzero elements of the $K$-sparse
signal $\x$, the support of $\x$ can be exactly recovered in $K$
iterations of OMP. The proposed conditions are sharp in terms of
$\delta_{K+1}$ for both types of noises, and the conditions on the
minimum magnitude of nonzero elements of $\x$ are weaker than
existing ones. We have also proposed worst-case necessary conditions for the
exact support recovery of $\x$ characterized by the minimum
magnitude of the nonzero elements of $\x$, under both $\ell_2$ and
$\ell_{\infty}$ bounded noises.

\section*{Acknowledgment}
We are  grateful to the editor and the anonymous referees for their valuable and
thoughtful suggestions that greatly improve the quality of this
paper.

\appendices

\section{Proof of Lemma~\ref{l:main}}
\label{ss:lemma1}

Before proving Lemma~\ref{l:main}, we introduce the following three useful lemmas, which were respectively proposed
in~\cite{CanT05},~\cite{NeeT09} and~\cite{SheLPL14}.
\begin{lemma}
\label{l:monot}
If $\A$ satisfies the RIP of orders $k_1$ and $k_2$ with $k_1<k_2$, then
$
\delta_{k_1}\leq \delta_{k_2}.
$
\end{lemma}

\begin{lemma}
\label{l:AtRIP}
Let $\A\in \mathbb{R}^{m\times n}$ satisfy the RIP of order $k$ and $S\subset\{1,2,\ldots, n\}$ with $|S|\leq k$, then for any $\x \in \mathbb{R}^m$,
\[
\|\A^T_S\x\|_2^2\leq(1+\delta_k)\|\x\|_2^2.
\]
\end{lemma}

\begin{lemma}
\label{l:orthogonalcomp}
Let sets $S_1,S_2$ satisfy $|S_2\setminus S_1|\geq1$ and matrix $\A$ satisfy the RIP of order $|S_1\cup S_2|$, then
for any vector $\x \in \mathbb{R}^{|S_2\setminus S_1|}$,
\beqnn
(1-\delta_{|S_1\cup S_2|})\|\x\|_2^2\leq \|\P^{\bot}_{S_1}\A_{S_2\setminus S_1}\x\|_2^2\leq(1+\delta_{|S_1\cup S_2|})\|\x\|_2^2.
\eeqnn
\end{lemma}

{\em Proof of Lemma~\ref{l:main}}. Obviously, to show~\eqref{e:main}, it suffices to show for each $j\in \Omega^c$,
\begin{align}
\label{e:main1}
\,\;&\|\A_{\Omega\setminus S}^T\P^{\bot}_S\A_{\Omega\setminus S}\x_{\Omega\setminus S}\|_{\infty}
-|\A_j^T\P^{\bot}_S\A_{\Omega\setminus S}\x_{\Omega\setminus S}|\nonumber\\
\geq & \frac{(1-\sqrt {|\Omega|-|S|+1 }\delta_{|\Omega|+1})\|\x_{\Omega\setminus S}\|_2}{\sqrt{|\Omega|-|S|}}.
\end{align}

Since $S$ is a proper subset of $\Omega$, $\|\x_{\Omega\setminus S}\|_1\neq0$. Hence
\begin{align*}
& \hspace{-4mm}\|\A_{\Omega\setminus S}^T\P^{\bot}_S\A_{\Omega\setminus S}\x_{\Omega\setminus S}\|_{\infty}\\
=&\frac{\sum_{\ell\in \Omega\setminus S}|x_{\ell}|}{\|\x_{\Omega\setminus S}\|_1}
\|\A_{\Omega\setminus S}^T\P^{\bot}_S\A_{\Omega\setminus S}\x_{\Omega\setminus S}\|_{\infty}\\
\overset{(a)}{\geq}&\frac{1}{\sqrt{|\Omega|-|S|}\|\x_{\Omega\setminus S}\|_2}(\sum_{\ell\in \Omega\setminus S}|x_{\ell}|)
\|\A_{\Omega\setminus S}^T\P^{\bot}_S\A_{\Omega\setminus S}\x_{\Omega\setminus S}\|_{\infty}\\
\geq&\frac{1}{\sqrt{|\Omega|-|S|}\|\x_{\Omega\setminus S}\|_2}\sum_{\ell\in \Omega\setminus S}
\big(x_{\ell}\A_{\ell}^T\P^{\bot}_S\A_{\Omega\setminus S}\x_{\Omega\setminus S}\big)\\
=&\frac{1}{\sqrt{|\Omega|-|S|}\|\x_{\Omega\setminus S}\|_2}
\big(\sum_{\ell\in \Omega\setminus S}x_{\ell}\A_{\ell}\big)^T\P^{\bot}_S\A_{\Omega\setminus S}\x_{\Omega\setminus S}\\
=&\frac{1}{\sqrt{|\Omega|-|S|}\|\x_{\Omega\setminus S}\|_2}
\big(\A_{\Omega\setminus S}\x_{\Omega\setminus S}\big)^T\P^{\bot}_S\A_{\Omega\setminus S}\x_{\Omega\setminus S}\\
\overset{(b)}{=}&\frac{1}{\sqrt{|\Omega|-|S|}
\|\x_{\Omega\setminus S}\|_2}\big(\P^{\bot}_S\A_{\Omega\setminus S}\x_{\Omega\setminus S}\big)^T\P^{\bot}_S\A_{\Omega\setminus S}\x_{\Omega\setminus S}\\
=&\frac{1}{\sqrt{|\Omega|-|S|}\|\x_{\Omega\setminus S}\|_2}\|\P^{\bot}_S\A_{\Omega\setminus S}\x_{\Omega\setminus S}\|_2^2,
\end{align*}
where (a) follows from $|\text{supp}(\x_{\Omega\setminus S})|=|\Omega|-|S|$ and the fact that
$\|\x\|_1\leq \sqrt{|\text{supp}(\x)|}\|\x\|_2$ for all $x\in \mathbb{R}^n$
(For more details, see, e.g., \cite[p.517]{FouR13}.  Note that this inequality itself can  be derived from the Cauchy-Schwarz inequality),
and (b) is because $\P^{\bot}_S$ is an orthogonal projector which has the idempotent and symmetry properties, i.e.,
\begin{align}
\label{e:orthcom}
(\P^{\bot}_S)^T\P^{\bot}_S=\P^{\bot}_S\P^{\bot}_S=\P^{\bot}_S.
\end{align}
Thus,
\begin{align}
\label{e:Ax}
&\hspace{-6mm} \|\P^{\bot}_S\A_{\Omega\setminus S}\x_{\Omega\setminus S}\|_2^2\nonumber\\
\leq&\sqrt{|\Omega|-|S|}\|\x_{\Omega\setminus S}\|_2\|\A_{\Omega\setminus S}^T\P^{\bot}_S\A_{\Omega\setminus S}\x_{\Omega\setminus S}\|_{\infty}.
\end{align}

Let
\beq
\label{e:alpha}
\alpha=-\frac{\sqrt{|\Omega|-|S|+1}-1}{\sqrt{|\Omega|-|S|}}.
\eeq
Then, by some simple calculations, we obtain
\beq
\label{e:alphaproperty}
\frac{2\alpha}{1-\alpha^2}=-\sqrt{|\Omega|-|S|}, \,\;\frac{1+\alpha^2}{1-\alpha^2}=\sqrt{|\Omega|-|S|+1}.
\eeq

To simplify the notation, for given $j\in \Omega^c$, we define
\begin{align}
\label{e:B}
\B=&\P^{\bot}_S
\bmx
\A_{\Omega\setminus S}&\A_{j}
\emx
,\\
\u=&
\bmx
\x_{\Omega\setminus S}^T&0
\emx^T\in \mathbb{R}^{|\Omega\setminus S|+1}, \nonumber\\
\w=&
\bmx
\0^T&\alpha t\|\x_{\Omega\setminus S}\|_2
\emx^T\in \mathbb{R}^{|\Omega\setminus S|+1},
\label{e:w}
\end{align}
where
\beq
\label{e:t}
t=
 \begin{cases}
      1 & \mbox{if }\A^T_{j}\P^{\bot}_S\A_{\Omega\setminus S}\x_{\Omega\setminus S}\geq 0 \\
      -1 & \mbox{if }\A^T_{j}\P^{\bot}_S\A_{\Omega\setminus S}\x_{\Omega\setminus S}< 0
   \end{cases}.
\eeq
Then,
\begin{align}
\label{e:AB}
\P^{\bot}_S\A_{\Omega\setminus S}\x_{\Omega\setminus S}=\B\u,
\end{align}
and
\begin{align}
\label{e:uw}
\|\u+\w\|_2^2&=(1+\alpha^2)\|\x_{\Omega\setminus S}\|_2^2, \\
\label{e:alphauw}
\|\alpha^2\u-\w\|_2^2&=\alpha^2(1+\alpha^2)\|\x_{\Omega\setminus S}\|_2^2.
\end{align}
Thus
\begin{align*}
 \w^T\B^T\B\u
\overset{(a)}{=}&\alpha t\|\x_{\Omega\setminus S}\|_2\A^T_{j}(\P^{\bot}_S)^T\P^{\bot}_S\A_{\Omega\setminus S}\x_{\Omega\setminus S}\\
\overset{(b)}{=}&\alpha t\|\x_{\Omega\setminus S}\|_2\A^T_{j}\P^{\bot}_S\A_{\Omega\setminus S}\x_{\Omega\setminus S}\\
\overset{(c)}{=}&\alpha\|\x_{\Omega\setminus S}\|_2|\A^T_{j}\P^{\bot}_S\A_{\Omega\setminus S}\x_{\Omega\setminus S}|,
\end{align*}
where (a) follows from~\eqref{e:B}, \eqref{e:w} and~\eqref{e:AB}, (b) follows from~\eqref{e:orthcom},
and (c) is from \eqref{e:t}. Therefore, for $j\in \Omega^c$, we have
\begin{align*}
&\hspace{-6mm}\|\B(\u+\w)\|_2^2\\
=&\|\B\u\|_2^2+\|\B\w\|_2^2+2\w^T\B^T\B\u\\
=&\|\B\u\|_2^2+\|\B\w\|_2^2+2\alpha\|\x_{\Omega\setminus S}\|_2|\A^T_{j}\P^{\bot}_S\A_{\Omega\setminus S}\x_{\Omega\setminus S}|
\end{align*}
and
\begin{align*}
 &\hspace{-4mm} \|\B(\alpha^2\u-\w)\|_2^2\\
=&\alpha^4\|\B\u\|_2^2+\|\B\w\|_2^2-2\alpha^3\|\x_{\Omega\setminus S}\|_2|\A^T_{j}\P^{\bot}_S\A_{\Omega\setminus S}\x_{\Omega\setminus S}|.
\end{align*}
By the aforementioned equations, we have
\begin{align}
&\hspace{-4mm} \|\B(\u+\w)\|_2^2-\|\B(\alpha^2\u-\w)\|_2^2 \nonumber \\
=&(1-\alpha^4)\|\B\u\|_2^2\nonumber \\
\quad&+2\alpha(1+\alpha^2)\|\x_{\Omega\setminus S}\|_2|\A^T_{j}\P^{\bot}_S\A_{\Omega\setminus S}\x_{\Omega\setminus S}|  \nonumber\\
=&(1-\alpha^4) \left(\|\B\u\|_2^2
+\frac{2\alpha}{1-\alpha^2}\|\x_{\Omega\setminus S}\|_2|\A^T_{j}\P^{\bot}_S\A_{\Omega\setminus S}\x_{\Omega\setminus S}|\right) \nonumber\\
=&(1-\alpha^4)\nonumber \\
\quad&\times(\|\B\u\|_2^2-\sqrt{|\Omega|-|S|}\|\x_{\Omega\setminus S}\|_2|\A^T_{j}\P^{\bot}_S\A_{\Omega\setminus S}\x_{\Omega\setminus S}|),
\label{e:transf11}
\end{align}
where the last equality follows from the first equality in~\eqref{e:alphaproperty}.

It is not hard to check that
\begin{align}
&\hspace{-6mm}\|\B(\u+\w)\|_2^2-\|\B(\alpha^2\u-\w)\|_2^2 \nonumber \\
\overset{(a)}{\geq}&(1-\delta_{|\Omega|+1})\|(\u+\w)\|_2^2
-(1+\delta_{|\Omega|+1})\|(\alpha^2\u-\w)\|_2^2\nonumber \\
\overset{(b)}{=}&(1-\delta_{|\Omega|+1})(1+\alpha^2)\|\x_{\Omega\setminus S}\|_2^2\nonumber \\
&-(1+\delta_{|\Omega|+1})\alpha^2(1+\alpha^2)\|\x_{\Omega\setminus S}\|_2^2\nonumber \\
=&(1+\alpha^2)\|\x_{\Omega\setminus S}\|_2^2\big((1-\delta_{|\Omega|+1})-(1+\delta_{|\Omega|+1})\alpha^2\big)\nonumber \\
=&(1+\alpha^2)\|\x_{\Omega\setminus S}\|_2^2\big((1-\alpha^2)-\delta_{|\Omega|+1}(1+\alpha^2)\big)\nonumber \\
=&(1-\alpha^4)\|\x_{\Omega\setminus S}\|_2^2\left(1-\frac{1+\alpha^2}{1-\alpha^2}\delta_{|\Omega|+1}\right)\nonumber \\
\overset{(c)}{=}&(1-\alpha^4)\|\x_{\Omega\setminus S}\|_2^2\big(1-\sqrt{|\Omega|-|S|+1}\delta_{|\Omega|+1}\big),
\label{e:transf12}
\end{align}
where (a) follows from Lemma~\ref{l:orthogonalcomp} and~\eqref{e:B}, (b) is from~\eqref{e:uw} and~\eqref{e:alphauw},
and (c) follows from the second equality in~\eqref{e:alphaproperty}.

By~\eqref{e:AB},~\eqref{e:transf11},~\eqref{e:transf12} and the fact that $1-\alpha^4>0$, we have
\begin{align*}
&\hspace{-.5mm}\|\P^{\bot}_S\A_{\Omega\setminus S}\x_{\Omega\setminus S}\|_2^2 \hspace{-.5mm} - \hspace{-.5mm} \sqrt{|\Omega| \hspace{-.5mm} - \hspace{-.5mm} |S|}\|\x_{\Omega\setminus S}\|_2|\A^T_{j}\P^{\bot}_S\A_{\Omega\setminus S}\x_{\Omega\setminus S}|\\
&= \|\B\u\|_2^2-\sqrt{|\Omega|-|S|}\|\x_{\Omega\setminus S}\|_2|\A^T_{j}\P^{\bot}_S\A_{\Omega\setminus S}\x_{\Omega\setminus S}|\\
&\geq \|\x_{\Omega\setminus S}\|_2^2\big(1-\sqrt{|\Omega|-|S|+1}\delta_{|\Omega|+1}\big).
\end{align*}
Combining the aforementioned equation with~\eqref{e:Ax}, we obtain
\begin{align*}
\,&\sqrt{|\Omega|-|S|}\|\x_{\Omega\setminus S}\|_2\\
&\quad\times\big(\|\A_{\Omega\setminus S}^T\P^{\bot}_S\A_{\Omega\setminus S}\x_{\Omega\setminus S}\|_{\infty}
-|\A_j^T\P^{\bot}_S\A_{\Omega\setminus S}\x_{\Omega\setminus S}|\big)\\
& ~~~\geq\|\x_{\Omega\setminus S}\|_2^2\big(1-\sqrt{|\Omega|-|S|+1}\delta_{|\Omega|+1}\big).
\end{align*}
Therefore,~\eqref{e:main1} holds, which establishes the lemma.
\ \ $\Box$

\section{Proof of Theorem~\ref{t:l2}}
\label{ss:l2}

{\em Proof}. Our proof consists of two steps. First, we show that OMP makes correct selection at each iteration. Then, we prove that it performs exactly $|\Omega|$ iterations.

We prove the first step by induction. Suppose that the OMP algorithm selects correct indices in the first $k-1$ iterations,
i.e., $S_{k-1}\subseteq \Omega$.
Then, we need to show that the OMP algorithm also selects a correct index at the $k$-th iteration, i.e., showing that $s^k\in \Omega$ (see Algorithm~\ref{a:OMP}).
Here, we assume $1\leq k\leq |\Omega|$. Thus, the proof for the first selection corresponds to the case of $k=1$. Clearly the induction hypothesis $S_{k-1}\subseteq \Omega$ holds for this case since $S_0=\emptyset$.

By line 2 of Algorithm~\ref{a:OMP}, to show $s^k\in \Omega$, it is equivalent to show that
\beq
\label{e:l2condoriginal}
\max_{i\in \Omega}|\langle \rr^{k-1},\A_i\rangle|
> \max_{j\in \Omega^c}|\langle \rr^{k-1},\A_j\rangle|.
\eeq
In the following, we simplify \eqref{e:l2condoriginal}.
Since the minimum eigenvalue of $\A^T_{S_{k-1}}\A_{S_{k-1}}$ is lower bounded by
\[
1-\delta_{|S_{k-1}|}\geq1-\delta_{|\Omega|}>0,
\]
$\A^T_{S_{k-1}}\A_{S_{k-1}}$ is invertible.
Thus, by line 4 of Algorithm~\ref{a:OMP}, we have
\begin{align}
\label{e:xtk-1}
\hat{\x}_{S_{k-1}}=(\A_{S_{k-1}}^T\A_{S_{k-1}})^{-1}\A_{S_{k-1}}^T\y.
\end{align}
Then, it follows from line 5 of Algorithm~\ref{a:OMP} and~\eqref{e:xtk-1} that
\begin{align}
\label{e:rk-1}
\rr^{k-1}&=\y-\A_{S_{k-1}}\hat{\x}_{S_{k-1}}\nonumber \\
&=\big(\I-\A_{S_{k-1}}(\A_{S_{k-1}}^T\A_{S_{k-1}})^{-1}\A_{S_{k-1}}^T\big)\y \nonumber \\
&\overset{(a)}{=}\P^{\perp}_{S_{k-1}}(\A\x+\v) \nonumber \\
&\overset{(b)}{=}\P^{\perp}_{S_{k-1}}(\A_{\Omega}\x_{\Omega}+\v) \nonumber \\
&\overset{(c)}{=}\P^{\perp}_{S_{k-1}}(\A_{S_{k-1}}\x_{S_{k-1}}+\A_{\Omega\setminus S_{k-1}}\x_{\Omega\setminus S_{k-1}}+\v) \nonumber \\
&\overset{(d)}{=}\P^{\perp}_{S_{k-1}}\A_{\Omega\setminus S_{k-1}}\x_{\Omega\setminus S_{k-1}}+\P^{\perp}_{S_{k-1}}\v,
\end{align}
where (a) follows from the definition of $\P^{\perp}_{S_{k-1}}$,
(b) is due to the fact that $\Omega=\text{supp}(\x)$,
(c) is from the induction assumption that $S_{k-1}\subseteq \Omega$,
and (d) follows from
\beq
\label{e:projpropty}
\P^{\perp}_{S_{k-1}}\A_{S_{k-1}}=\0.
\eeq

Thus, by \eqref{e:rk-1} and \eqref{e:projpropty}, for $i\in S_{k-1}$, we have
\[
\langle \rr^{k-1},\A_i\rangle=\A_i^T\rr^{k-1}=0.
\]
Therefore, to show \eqref{e:l2condoriginal}, it is equivalent to show
\beq
\label{e:l2cond}
\max_{i\in \Omega\setminus S_{k-1}}|\langle \rr^{k-1},\A_i\rangle|
> \max_{j\in \Omega^c}|\langle \rr^{k-1},\A_j\rangle|.
\eeq

In the following, we will use \eqref{e:rk-1} to rewrite \eqref{e:l2cond}.
By~\eqref{e:rk-1} and the (reverse) triangle inequality, we obtain
\begin{align}
\label{e:left}
&\hspace{-4mm}\max_{i\in \Omega\setminus S_{k-1}}|\langle \rr^{k-1},\A_i\rangle| \nonumber\\
& =\|\A^T_{\Omega\setminus S_{k-1}}\big(\P^{\perp}_{S_{k-1}}\A_{\Omega\setminus S_{k-1}}\x_{\Omega\setminus S_{k-1}}+\P^{\perp}_{S_{k-1}}\v\big)\|_{\infty}
\nonumber\\
& \geq \|\A^T_{\Omega\setminus S_{k-1}}\P^{\perp}_{S_{k-1}}\A_{\Omega\setminus S_{k-1}}\x_{\Omega\setminus S_{k-1}}\|_{\infty} \nonumber \\
& ~~~~-\|\A^T_{\Omega\setminus S_{k-1}}\P^{\perp}_{S_{k-1}}\v\|_{\infty},
\end{align}
and
\begin{align}
\label{e:right}
&\hspace{-4mm} \max_{j\in \Omega^c}|\langle \rr^{k-1},\A_j\rangle|\nonumber\\
& =\|\A^T_{\Omega^c}\big(\P^{\perp}_{S_{k-1}}\A_{\Omega\setminus S_{k-1}}\x_{\Omega\setminus S_{k-1}}+\P^{\perp}_{S_{k-1}}\v\big)\|_{\infty}
\nonumber\\
& \leq \|\A^T_{\Omega^c}\P^{\perp}_{S_{k-1}}\A_{\Omega\setminus S_{k-1}}\x_{\Omega\setminus S_{k-1}}\|_{\infty}
+\|\A^T_{\Omega^c}\P^{\perp}_{S_{k-1}}\v\|_{\infty}.
\end{align}
Therefore, by~\eqref{e:left} and~\eqref{e:right}, to show~\eqref{e:l2cond}, it suffices to show
\begin{align}
&\hspace{-4mm}\|\A^T_{\Omega\setminus S_{k-1}}\P^{\perp}_{S_{k-1}}\A_{\Omega\setminus S_{k-1}}\x_{\Omega\setminus S_{k-1}}\|_{\infty}\nonumber\\
&-\|\A^T_{\Omega^c}\P^{\perp}_{S_{k-1}}\A_{\Omega\setminus S_{k-1}}\x_{\Omega\setminus S_{k-1}}\|_{\infty} \nonumber\\
&> \|\A^T_{\Omega\setminus S_{k-1}}\P^{\perp}_{S_{k-1}}\v\|_{\infty}+\|\A^T_{\Omega^c}\P^{\perp}_{S_{k-1}}\v\|_{\infty}
\label{e:l2cond1}.
\end{align}

We next give a lower bound on the left-hand side of~\eqref{e:l2cond1}.
By the induction assumption $S_{k-1}\subseteq \Omega$, we have
\begin{align}
\label{e:suppxk-1}
|\text{supp}(\x_{\Omega\setminus S_{k-1}})|=|\Omega|+1-k.
\end{align}
Thus,
\begin{align}
\label{e:minxk-1}
\|\x_{\Omega\setminus S_{k-1}}\|_2&\geq \sqrt {|\Omega|+1-k }\min_{i\in\Omega\setminus S_{k-1}}|x_i| \nonumber\\
&\geq \sqrt {|\Omega|+1-k }\min_{i\in\Omega}|x_i|.
\end{align}

Since $S_{k-1}\subseteq \Omega$ and $|S_{k-1}|=k-1$, by Lemma~\ref{l:main}, we have
\begin{align}
&\hspace{-4mm}\|\A^T_{\Omega\setminus S_{k-1}}\P^{\perp}_{S_{k-1}}\A_{\Omega\setminus S_{k-1}}\x_{\Omega\setminus S_{k-1}}\|_{\infty}\nonumber\\
&-\|\A^T_{\Omega^c}\P^{\perp}_{S_{k-1}}\A_{\Omega\setminus S_{k-1}}\x_{\Omega\setminus S_{k-1}}\|_{\infty} \nonumber\\
& \geq \frac{(1-\sqrt {|\Omega|-k+2 }\delta_{|\Omega|+1})\|\x_{\Omega\setminus S_{k-1}}\|_2}{\sqrt{|\Omega|+1-k}}\nonumber\\
& \overset{(a)}{\geq} \frac{(1-\sqrt {K+1 }\delta_{|\Omega|+1})\|\x_{\Omega\setminus S_{k-1}}\|_2}{\sqrt{|\Omega|+1-k}}\nonumber\\
& \overset{(b)}{\geq} \frac{(1-\sqrt {K+1 }\delta_{K+1})\|\x_{\Omega\setminus S_{k-1}}\|_2}{\sqrt{|\Omega|+1-k}}\nonumber\\
& \overset{(c)}{\geq} (1-\sqrt {K+1 }\delta_{K+1})\min_{i\in\Omega}|x_i|,
\label{e:l2cond1left}
\end{align}
where (a) is because $k\geq1$ and $\x$ is $K$-sparse (i.e., $|\Omega|\leq K$), (b) follows from Lemma~\ref{l:monot}, and (c) is from~\eqref{e:delta} and~\eqref{e:minxk-1}.

We next give an upper bound on the right-hand side of~\eqref{e:l2cond1}.
Obviously, there exist $i_0\in\Omega\setminus S_{k-1}$ and $j_0\in\Omega^c$ such that
\begin{align}
\label{e:l2v1}
\|\A^T_{\Omega\setminus S_{k-1}}\P^{\perp}_{S_{k-1}}\v\|_{\infty}&=|\A^T_{i_0}\P^{\perp}_{S_{k-1}}\v|,\\
\label{e:l2v2}
\|\A^T_{\Omega^c}\P^{\perp}_{S_{k-1}}\v\|_{\infty}&=|\A^T_{j_0}\P^{\perp}_{S_{k-1}}\v|.
\end{align}
Hence,
\begin{align}
\label{e:l2cond1right}
\,&\|\A^T_{\Omega\setminus S_{k-1}}\P^{\perp}_{S_{k-1}}\v\|_{\infty}+\|\A^T_{\Omega^c}\P^{\perp}_{S_{k-1}}\v\|_{\infty}\nonumber \\
=&|\A^T_{i_0}\P^{\perp}_{S_{k-1}}\v|+|\A^T_{j_0}\P^{\perp}_{S_{k-1}}\v|\nonumber \\
=&\|\A^T_{i_0\cup j_0}\P^{\perp}_{S_{k-1}}\v\|_1\nonumber \\
\overset{(a)}{\leq}&\sqrt{2}\|\A^T_{i_0\cup j_0}\P^{\perp}_{S_{k-1}}\v\|_2\nonumber \\
\overset{(b)}{\leq}& \sqrt{2(1+\delta_{K+1})}\|\P^{\perp}_{S_{k-1}}\v\|_2\nonumber \\
\overset{(c)}{\leq}& \sqrt{2(1+\delta_{K+1})}\epsilon,
\end{align}
where (a) is due to that $\A^T_{i_0\cup j_0}\P^{\perp}_{S_{k-1}}\v$ is a $2\times1$ vector and that
$\|\x\|_1\leq \sqrt{|\text{supp}(\x)|}\|\x\|_2$ for all $x\in \mathbb{R}^n$
(For more details, see e.g., \cite[p.517]{FouR13}. Note that this inequality itself can  be derived from the Cauchy-Schwarz inequality),
(b) follows from Lemma~\ref{l:AtRIP} and (c) is because
\beq
\label{e:orthcompv}
\|\P^{\perp}_{S_{k-1}}\v\|_2\leq \|\v\|_2\leq\epsilon.
\eeq

From~\eqref{e:l2cond1left} and~\eqref{e:l2cond1right},~\eqref{e:l2cond1} (or equivalently~\eqref{e:l2cond}) can be guaranteed by
\begin{align*}
(1-\sqrt {K+1 }\delta_{K+1})\min_{i\in\Omega}|x_i|> \sqrt{2(1+\delta_{K+1})}\epsilon,
\end{align*}
i.e.,
\beqnn
\min_{i\in\Omega}|x_i|> \frac{\sqrt{2(1+\delta_{K+1})}\epsilon}{1-\sqrt {K+1 }\delta_{K+1}}.
\eeqnn
Furthermore, by~\eqref{e:delta}, we have $\sqrt{1+\delta_{K+1}}<\sqrt{2}$.
Thus, if~\eqref{e:l2} holds, then the OMP algorithm selects a correct index in each iteration.

Now, what remains to show is that the OMP algorithm performs exact $|\Omega|$ iterations,
which is equivalent to show that $ \|\rr^k\|_2>\epsilon$ for $1\leq k<|\Omega|$ and $ \|\rr^{|\Omega|}\|_2\leq\epsilon$.

Since OMP selects a correct index at each iteration under~\eqref{e:l2},
by the (reverse) triangle inequality and~\eqref{e:rk-1}, for $1\leq k<|\Omega|$, we have
\begin{align}
\|\rr^k\|_2&= \|\P^{\perp}_{S_{k}}\A_{\Omega\setminus S_{k}}\x_{\Omega\setminus S_{k}}+\P^{\perp}_{S_{k}}\v\|_2\nonumber\\
&\geq \|\P^{\perp}_{S_{k}}\A_{\Omega\setminus S_{k}}\x_{\Omega\setminus S_{k}}\|_2-\|\P^{\perp}_{S_{k}}\v\|_2\nonumber\\
&\overset{(a)}{\geq}\|\P^{\perp}_{S_{k}}\A_{\Omega\setminus S_{k}}\x_{\Omega\setminus S_{k}}\|_2-\epsilon\nonumber\\
&\overset{(b)}{\geq}\sqrt{1-\delta_{|\Omega|}}\|\x_{\Omega\setminus\S_k}\|_2-\epsilon\nonumber\\
&\overset{(c)}{\geq}\sqrt{1-\delta_{K+1}}\sqrt{|\Omega|-k}\min_{i\in\Omega}|x_i|-\epsilon\nonumber\\
&\geq\sqrt{1-\delta_{K+1}}\min_{i\in\Omega}|x_i|-\epsilon,
\label{e:rklbd}
\end{align}
where (a) is from~\eqref{e:orthcompv}, (b) is from Lemma~\ref{l:orthogonalcomp},
and (c) follows from Lemma~\ref{l:monot} and~\eqref{e:minxk-1}.
Therefore, if
\beq
\label{e:earlycondl2}
\min_{i\in\Omega}|x_i|> \frac{2\epsilon}{\sqrt {1-\delta_{K+1}}},
\eeq
then $\|\rr^k\|_2>\epsilon$ for each $1\leq k< \Omega$.

By some simple calculations, we can show that
\beq
\label{e:condl22}
\frac{2\epsilon}{1-\sqrt {K+1 }\delta_{K+1}}\geq \frac{2\epsilon}{\sqrt {1-\delta_{K+1}}}.
\eeq
Indeed, by the fact that $0<1-\delta_{K+1}<1$, we have
\beqnn
1-\sqrt {K+1 }\delta_{K+1}\leq1-\delta_{K+1}\leq\sqrt {1-\delta_{K+1}}.
\eeqnn
Thus,~\eqref{e:condl22} holds.

Therefore, by~\eqref{e:earlycondl2} and~\eqref{e:condl22}, if~\eqref{e:l2} holds, $\|\rr^k\|_2>\epsilon$ for each $1\leq k< \Omega$,
i.e., the OMP algorithm does not terminate before the $|\Omega|$-th iteration.

Similarly, by~\eqref{e:rk-1},
\begin{align}
\label{e:rKbd}
\|\rr^{|\Omega|}\|_2&= \|\P^{\perp}_{S_{|\Omega|}}\A_{\Omega\setminus S_{|\Omega|}}
\x_{\Omega\setminus S_{|\Omega|}}+\P^{\perp}_{S_{|\Omega|}}\v\|_2\nonumber\\
&\overset{(a)}{=}\|\P^{\perp}_{S_{|\Omega|}}\v\|_2
\overset{(b)}{\leq}\epsilon,
\end{align}
where (a) is because $S_{|\Omega|}=|\Omega|$ and (b) follows from~\eqref{e:orthcompv}.
So, by the stopping condition, the OMP algorithm terminates after the $|\Omega|$-th iteration.
Therefore, the OMP algorithm performs $|\Omega|$ iterations. This completes the proof.\ \ $\Box$

\section{Proof of Theorem~\ref{t:linf}}
\label{ss:linf}
{\em Proof}. Similar to the proof of Theorem~\ref{t:l2}, we need to prove that the OMP algorithm selects correct indexes at all iterations
and it performs exactly $|\Omega|$ iterations.

We first prove the first part. By the proof of Theorem~\ref{t:l2}, we only need to prove that~\eqref{e:l2cond1} holds.
As the noise vector satisfies a different constraint, we need to give a new upper bound on the right-hand side of~\eqref{e:l2cond1}.
To do this, we first use the method used in the proof of ~\cite[Theorem 5]{CaiW11} to give an upper bound on
$\|\P_{S_{k-1}}\v\|_{2}$ and then use~\eqref{e:l2v1} and~\eqref{e:l2v2} to given an upper bound on
$$
\|\A^T_{\Omega\setminus S_{k-1}}\P^{\perp}_{S_{k-1}}\v\|_{\infty}+\|\A^T_{\Omega^c}\P^{\perp}_{S_{k-1}}\v\|_{\infty}.
$$

Let $\lambda$ denote the largest singular value of $(\A_{S_{k-1}}^T\A_{S_{k-1}})^{-1}$. Then $\lambda$ equals to the reciprocal
of the smallest singular value of $\A_{S_{k-1}}^T\A_{S_{k-1}}$.
Since $\A$ satisfies the RIP of order $K+1$ with $\delta_{K+1}$,
the smallest singular value of $\A_{S_{k-1}}^T\A_{S_{k-1}}$ cannot be smaller than $1-\delta_{K+1}$.
Thus, $\lambda\leq 1/(1-\delta_{K+1})$. Therefore,
\begin{align}
\label{e:pv}
\|\P_{S_{k-1}}\v\|_{2}^2&=\v^T\P_{S_{k-1}}^T\P_{S_{k-1}}\v=\v^T\P_{S_{k-1}}\v\nonumber \\
&\overset{(a)}{=}\v^T\A_{S_{k-1}}(\A_{S_{k-1}}^T\A_{S_{k-1}})^{-1}\A_{S_{k-1}}^T\v\nonumber \\
&\overset{(b)}{\leq}\lambda \|\A_{S_{k-1}}^T\v\|_2^2\nonumber \\
&\leq \frac{1}{1-\delta_{K+1}} \|\A_{S_{k-1}}^T\v\|_{2}^2\nonumber \\
&\overset{(c)}{\leq} \frac{K-1}{1-\delta_{K+1}} \|\A_{S_{k-1}}^T\v\|_{\infty}^2\nonumber \\
&\overset{(d)}{\leq}\frac{(K-1)\epsilon^2}{1-\delta_{K+1}}
<\frac{K\epsilon^2}{1-\delta_{K+1}},
\end{align}
where (a) follows from the definition of $\P_{S_{k-1}}$,
(b) is from the assumption that $\lambda$ is the largest singular value of $(\A_{S_{k-1}}^T\A_{S_{k-1}})^{-1}$,
(c) is because $|S_{k-1}|=k-1\leq K-1$,
and (d) follows from $\|\A^T\v\|_{\infty}\leq \epsilon$.

By~\eqref{e:l2v1},~\eqref{e:l2v2} and the triangular inequality, we have
\begin{align}
\label{e:linfcond1right}
&\hspace{-8mm}\|\A^T_{\Omega\setminus S_{k-1}}\P^{\perp}_{S_{k-1}}\v\|_{\infty}+\|\A^T_{\Omega^c}\P^{\perp}_{S_{k-1}}\v\|_{\infty}\nonumber \\
=&|\A^T_{i_0}\P^{\perp}_{S_{k-1}}\v|+|\A^T_{j_0}\P^{\perp}_{S_{k-1}}\v|\nonumber \\
=&\|\A^T_{i_0\cup j_0}\P^{\perp}_{S_{k-1}}\v\|_1\nonumber \\
\overset{(a)}{\leq}&\sqrt{2}\|\A^T_{i_0\cup j_0}\P^{\perp}_{S_{k-1}}\v\|_2\nonumber \\
=&\sqrt{2}\|\A^T_{i_0\cup j_0}(\I-\P_{S_{k-1}})\v\|_2\nonumber \\
\leq&\sqrt{2}\|\A^T_{i_0\cup j_0}\v\|_2+\sqrt{2}\big\|\A^T_{i_0\cup j_0}\P_{S_{k-1}}\v\|_2\nonumber \\
\overset{(b)}{\leq}&2\|\A^T_{i_0\cup j_0}\v\|_{\infty}+
\sqrt{2(1+\delta_{2})}\big\|\P_{S_{k-1}}\v\|_2\nonumber \\
\overset{(c)}{\leq}&2\epsilon+
\sqrt{\frac{1+\delta_{2}}{1-\delta_{K+1}}
}\sqrt{2K}\epsilon,
\end{align}
where (a) is due to the fact that $\A^T_{i_0\cup j_0}\P^{\perp}_{S_{k-1}}\v$ is a $2\times 1$ vector and the Cauchy-Schwarz inequality,
(b) and (c) respectively follow from Lemma~\ref{l:AtRIP} and~\eqref{e:pv}.

Therefore, by~\eqref{e:l2cond1left} and~\eqref{e:linfcond1right}, if~\eqref{e:linf} holds, then~\eqref{e:l2cond1} holds,\footnote{Note that \eqref{e:l2cond1} still holds under the relaxed condition of \eqref{e:linf}.}
i.e., OMP selects correct indexes in all iterations if~\eqref{e:linf} holds.

Our next job is to prove that the OMP algorithm does not terminate before the $|\Omega|$-th iteration.
By the (reverse) triangular inequality and~\eqref{e:rk-1}, we have
\begin{align}
&\hspace{-6mm}\|\A^T\rr^k\|_{\infty}\nonumber\\
=& \|\A^T(\P^{\perp}_{S_{k}}\A_{\Omega\setminus S_{k}}\x_{\Omega\setminus S_{k}}+\P^{\perp}_{S_{k}}\v)\|_{\infty}\nonumber\\
\geq& \|\A^T_{\Omega\setminus S_{k}}(\P^{\perp}_{S_{k}}\A_{\Omega\setminus S_{k}}\x_{\Omega\setminus S_{k}}+\P^{\perp}_{S_{k}}\v)\|_{\infty}\nonumber\\
\geq& \|\A^T_{\Omega\setminus S_{k}}\P^{\perp}_{S_{k}}\A_{\Omega\setminus S_{k}}\x_{\Omega\setminus S_{k}}\|_{\infty}
-\|\A^T_{\Omega\setminus S_{k}}\P^{\perp}_{S_{k}}\v\|_{\infty}.
\label{e:rklbdinf}
\end{align}

In the following, we give a lower bound on
\[
\|\A^T_{\Omega\setminus S_{k}}\P^{\perp}_{S_{k}}\A_{\Omega\setminus S_{k}}\x_{\Omega\setminus S_{k}}\|_{\infty}.
\]
It is not hard to check that
\begin{align}
\label{e:linfkv1}
&\hspace{-6mm}\|\A^T_{\Omega\setminus S_{k}}\P^{\perp}_{S_{k}}\A_{\Omega\setminus S_{k}}\x_{\Omega\setminus S_{k}}\|_{\infty}\nonumber\\
\geq&\frac{1}{\sqrt{|\Omega|-k}}\|\A^T_{\Omega\setminus S_{k}}\P^{\perp}_{S_{k}}\A_{\Omega\setminus S_{k}}\x_{\Omega\setminus S_{k}}\|_{2}\nonumber\\
=&\frac{\|\x_{\Omega\setminus S_{k}}\|_2\|\A^T_{\Omega\setminus S_{k}}\P^{\perp}_{S_{k}}
\A_{\Omega\setminus S_{k}}\x_{\Omega\setminus S_{k}}\|_{2}}{\sqrt{|\Omega|-k}\|\x_{\Omega\setminus S_{k}}\|_2}\nonumber\\
\overset{(a)}{\geq}&\frac{|\x^T_{\Omega\setminus S_{k}}\A^T_{\Omega\setminus S_{k}}\P^{\perp}_{S_{k}}
\A_{\Omega\setminus S_{k}}\x_{\Omega\setminus S_{k}}|}{\sqrt{|\Omega|-k}\|\x_{\Omega\setminus S_{k}}\|_2}\nonumber\\
\overset{(b)}{=}&\frac{\|\P^{\perp}_{S_{k}}\A_{\Omega\setminus S_{k}}\x_{\Omega\setminus S_{k}}\|_2^2}{\sqrt{|\Omega|-k}\|\x_{\Omega\setminus S_{k}}\|_2}\nonumber\\
\overset{(c)}{\geq}&\frac{(1-\delta_{K})\|\x_{\Omega\setminus S_{k}}\|_2^2}{\sqrt{|\Omega|-k}\|\x_{\Omega\setminus S_{k}}\|_2}\nonumber\\
\geq&(1-\delta_{K+1})\frac{\|\x_{\Omega\setminus S_{k}}\|_2}{\sqrt{|\Omega|-k}}\nonumber\\
\geq& (1-\delta_{K+1})\min_{i\in\Omega}|x_i|,
\end{align}
where (a) follows from the Cauchy-Schwarz inequality, (b) is due to \eqref{e:orthcom},
(c) is from Lemma~\ref{l:orthogonalcomp},
and the last inequality is from \eqref{e:minxk-1}.

In the following, we give an upper bound on
\[
\|\A^T_{\Omega\setminus S_{k}}\P^{\perp}_{S_{k}}\v\|_{\infty}.
\]
Let $j_0\in \Omega\setminus S_{k}$ such that
$$
\|\A^T_{\Omega\setminus S_{k}}\P^{\perp}_{S_{k}}\v\|_{\infty}=|\A^T_{j_0}\P^{\perp}_{S_{k}}\v|.
$$
Then, by the triangular inequality, we obtain
\begin{align}
\label{e:linfv}
&\hspace{-6mm}\|\A^T_{\Omega\setminus S_{k}}\P^{\perp}_{S_{k}}\v\|_{\infty}=|\A^T_{j_0}(\I-\P_{S_{k}})\v|\nonumber \\
\leq& |\A^T_{j_0}\v|+|\A^T_{j_0}\P_{S_{k}}\v|\leq\epsilon+|\A^T_{j_0}\P_{S_{k}}\v|\nonumber \\
\leq& \left(1+
\sqrt{\frac{1+\delta_{1}}{1-\delta_{K+1}}
}\sqrt{K}\right)\epsilon,
\end{align}
where the last inequality follows from the Cauchy-Schwartz inequality, ~\eqref{e:pv} and Lemma~\ref{l:AtRIP}.

Therefore, for each $1\leq k< |\Omega|$, by~\eqref{e:linf} and~\eqref{e:rklbdinf}-\eqref{e:linfv}, we have
\begin{align*}
&\hspace{-6mm}\|\A^T\rr^k\|_{\infty}\\
\geq& (1-\delta_{K+1})\min_{i\in\Omega}|x_i|-\left(1+
\sqrt{\frac{1+\delta_{1}}{1-\delta_{K+1}}
}\sqrt{K}\right)\epsilon\\
>&
\frac{2(1-\delta_{K+1})}{1-\sqrt {K+1 }\delta_{K+1}}\left(1+\sqrt{\frac{1+\delta_{2}}{1-\delta_{K+1}}}\sqrt{K}\right)\epsilon
\\
&-
\left(1+\sqrt{\frac{1+\delta_{1}}{1-\delta_{K+1}}}\sqrt{K}\right)\epsilon
\\
\overset{(a)}{\geq}&
2\left(1+\sqrt{\frac{1+\delta_{2}}{1-\delta_{K+1}}}\sqrt{K}\right)\epsilon
-\left(1+\sqrt{\frac{1+\delta_{1}}{1-\delta_{K+1}}}\sqrt{K}\right)\epsilon
\\
\geq&
\left(1+\sqrt{\frac{1+\delta_{2}}{1-\delta_{K+1}}}\sqrt{K}\right)\epsilon,
\end{align*}
where (a) is because
\[
1-\delta_{K+1}\geq1-\sqrt{K+1}\delta_{K+1},
\]
and the last inequality follows from Lemma \ref{l:monot}.
So, by the stopping condition~\eqref{e:stopcinf}, the OMP algorithm does not terminate before the $|\Omega|$-th iteration.
\footnote{If $\A$ is column normalized, then $\delta_{1}=0$. Thus, under the relaxed condition of \eqref{e:linf},
we have
$
\|\A^T\rr^k\|_{\infty}\geq\left(1+\frac{\sqrt{K}}{\sqrt{1-\delta_{K+1}}}\right)\epsilon.
$
Thus, the OMP algorithm does not terminate before the $|\Omega|$-th iteration under the relaxed stopping condition.
}

Finally, we prove that OMP terminates after performing the $|\Omega|$-th iteration.
By~\eqref{e:rKbd}, we have
$$
\rr^{|\Omega|}=\P^{\perp}_{S_{|\Omega|}}\v.
$$
Thus, applying some techniques which are similar to that for deriving~\eqref{e:linfv}, we obtain
\begin{align*}
\|\A^T\rr^{|\Omega|}\|_{\infty}=&\|\A^T\P^{\perp}_{S_{|\Omega|}}\v\|_{\infty}\\
\leq& \left(1+\sqrt{\frac{1+\delta_{1}}{1-\delta_{K+1}}}\sqrt{K}\right)\epsilon,\\
\leq&
\left(1+\sqrt{\frac{1+\delta_{2}}{1-\delta_{K+1}}}\sqrt{K}\right)\epsilon,
\end{align*}
By the stopping condition~\eqref{e:stopcinf}, the OMP algorithm terminates after performing the $|\Omega|$-th iteration.
\footnote{If $\A$ is column normalized, then $\delta_{1}=0$. Thus,
$
\|\A^T\rr^{|\Omega|}\|_{\infty}\leq\left(1+\frac{\sqrt{K}}{\sqrt{1-\delta_{K+1}}}\right)\epsilon.
$
Thus, the OMP algorithm terminates after performing the $|\Omega|$-th iteration.
}
~~~~~~~~~~~~~~~~~$\Box$

\section{Proof of Theorem~\ref{t:l2nc}}
\label{ss:l2nc}

{\em Proof.}  To prove the theorem, it suffices to show that there exists a linear model of the form~\eqref{e:model},
where $\v$ satisfies $\|\v\|_2\leq \epsilon$,
$\A$ satisfies the RIP with $\delta_{K+1}=\delta$ for any given $\delta$ satisfying \eqref{e:deltanc},
and $\x$ is $K$-sparse and satisfies
\begin{eqnarray}
\label{e:l2nececounte}
\min_{i\in\Omega}|x_i|= \gamma
\end{eqnarray}
for some $\gamma$ satisfying
\beq
\label{e:gamma}
0<\gamma<\frac{\sqrt{1-\delta}\epsilon}{1-\sqrt {K+1 }\delta},
\eeq
such that the OMP algorithm fails to recover the support of $\x$ in $K$ iterations.

In the following, we construct such a linear model.

Let $\1_{K}$ be a $K$-dimensional column vector with all entries being 1, then
 there exist $\bxi_i\in \mathbb{R}^K, 1\leq i\leq K-1$, such that
\beqnn
\bmx
\bxi_1&\bxi_2&\ldots&\bxi_{K-1}&\frac{1}{\sqrt{K}}\1_{K}
\emx
\in \mathbb{R}^{K\times K}
\eeqnn
is an orthogonal matrix. Let
\beq
\label{e:U}
\U=\bmx
\bxi_1&\bxi_2&\ldots&\bxi_{K-1}&\frac{\1_{K}}{\sqrt{K(\beta^2+1)}}&\frac{\beta\1_{K}}{\sqrt{K(\beta^2+1)}}\\
0&0&\ldots&0&\frac{\beta}{\sqrt{\beta^2+1}}&-\frac{1}{\sqrt{\beta^2+1}}
\emx^T,
\eeq
where
\beqnn
\beta=\frac{\sqrt{K+1}-1}{\sqrt{K}}.
\eeqnn
Then, it is not hard to prove that $\U$ is also an orthogonal matrix.
Applying some techniques which are similar to that for deriving \eqref{e:alphaproperty},  we can show that
\beq
\label{e:beta}
\frac{1-\beta^2}{1+\beta^2}=\frac{1}{\sqrt{K+1}},\quad
\frac{2\beta}{1+\beta^2}=\frac{\sqrt{K}}{\sqrt{K+1}}.
\eeq

Let $\D\in\mathbb{R}^{(K+1)\times(K+1)}$ be the diagonal matrix with
\beq
\label{e:D}
d_{ii}=
\begin{cases}
\sqrt{1-\delta} &i=K\\
\sqrt{1+\delta} &i \neq K
\end{cases}
,
\eeq
and
\beq
\label{e:A}
\A=\D\U,
\eeq
then $\A^T\A=\U^T\D^2\U$.
It is not hard to see that $\A$ satisfies the RIP with $\delta_{K+1}=\delta$.
In fact, let $\V=\U^T$, then by the fact that $\U$ is orthogonal, we have
\[
\A^T\A=\U^T\D^2\U=\V\D^2\V^T=\V\D^2\V^{-1}.
\]
Thus, $\V\D^2\V^{-1}$ is a valid eigenvalue decomposition of $\A^T\A$, and $\D^2$ is the diagonal matrix containing the eigenvalues.
Therefore, the largest and smallest eigenvalues of $\A^T\A$ are respectively $1+\delta$ and $1-\delta$, which implies that $\delta_{K+1}=\delta$.

Let
\beq
\label{e:x}
\x=
\bmx
\gamma\1_{K}^T&0
\emx^T
\in \mathbb{R}^{K+1}
\eeq
for any $\gamma$ satisfying~\eqref{e:gamma} (recall that $\1_{K}$ is a $K$-dimensional column vector with all of its entries being 1).
Then $\x$ is $K$-sparse and satisfies~\eqref{e:l2nececounte}. By~\eqref{e:U} and the fact that $\bxi_i^T\1_{K}=0$ for $1\leq i\leq K-1$, we have
\beqnn
\U\x=\sqrt{\frac{K}{\beta^2+1}}\gamma
\bmx
\0_{K-1}^T&1&\beta
\emx^T.
\eeqnn
Thus,
\beq
\label{e:DUx}
\D^2\U\x=
\sqrt{\frac{K}{\beta^2+1}}\gamma
\bmx
\0_{K-1}^T&1-\delta&(1+\delta)\beta
\emx^T.
\eeq

Let
\beq
\label{e:v2}
\v=\D^{-1}\U
\bmx
\0_{K}^T& -\sqrt{1-\delta}\epsilon
\emx^T
\in \mathbb{R}^{K+1},
\eeq
then, by~\eqref{e:A} and the fact that $\U$ is orthogonal, we have
\beq
\label{e:Av}
\A^T\v=
\bmx
\0_{K}^T& -\sqrt{1-\delta}\epsilon
\emx^T
.
\eeq

In the following, we show that $\|\v\|_2\leq\epsilon$. By~\eqref{e:U} and~\eqref{e:v2}, we have
\[
\v=\D^{-1}\frac{\epsilon}{\sqrt{\beta^2+1}}
\bmx
\0_{K-1}^T& -\beta\sqrt{1-\delta}&\sqrt{1-\delta}
\emx^T.
\]
Thus, by~\eqref{e:D}, we obtain
$$
\v=\frac{\epsilon}{\sqrt{\beta^2+1}}
\bmx
\0_{K-1}^T& -\beta&\sqrt{\frac{1-\delta}{1+\delta}}
\emx^T,
$$
and hence $\|\v\|_2\leq\epsilon$.

Let $\e_i, 1\leq i\leq K,$ be the $i$-th column of the $(K+1)\times(K+1) $ identity matrix,
then,  for $1\leq i\leq K$, we have
\begin{align*}
\langle \y,\A_i\rangle
&=\langle \A\x+\v,\A_i\rangle=\e_i^T\A^T\A\x+\e_i^T\A^T\v\\
&\overset{(a)}{=}\e_i^T\U^T\D^2\U\x\\
&\overset{(b)}{=}\left(\frac{(1-\delta)+(1+\delta)\beta^2}{\beta^2+1}\right)\gamma\\
&=\left(1-\frac{1-\beta^2}{1+\beta^2}\delta\right)\gamma\\
&=\left(1-\frac{1}{\sqrt{K+1}}\delta\right)\gamma,
\end{align*}
where (a) follows from the fact that $\U$ is orthogonal, \eqref{e:A} and \eqref{e:Av},
(b) is due to \eqref{e:U} and \eqref{e:DUx},
and the last equality is from~\eqref{e:beta}. Thus, by \eqref{e:deltanc},
\beq
\label{e:countleftl2}
\max_{i\in\Omega} |\langle \y,\A_i\rangle|=\left(1-\frac{1}{\sqrt{K+1}}\delta\right)\gamma.
\eeq

Similarly, we have
\begin{align*}
\langle \y,\A_{K+1}\rangle
&=\e_{K+1}^T\U^T\D^2\U\x+\e_{K+1}^T\A^T\v\\
&=\frac{-2\beta}{\beta^2+1}\sqrt{K}\delta\gamma -\sqrt{1-\delta}\epsilon\\
&=\frac{-K}{\sqrt{K+1}}\delta\gamma -\sqrt{1-\delta}\epsilon,
\end{align*}
where the last equality is from~\eqref{e:beta}. Thus
\beq
\label{e:countrightl2}
\max_{j\in\Omega^c} |\langle \y,\A_j\rangle|=\frac{K}{\sqrt{K+1}}\delta\gamma +\sqrt{1-\delta}\epsilon.
\eeq

By~\eqref{e:gamma},~\eqref{e:countleftl2} and~\eqref{e:countrightl2}, we have
$$
\max_{i\in\Omega} |\langle \y,\A_i\rangle|< \max_{j\in\Omega^c} |\langle \y,\A_j\rangle|.
$$
Thus, by line 2 of Algorithm~\ref{a:OMP}, OMP chooses an index in $\Omega^c$ in the first iteration.
Therefore, OMP fails to recover $\Omega$ in $K$ iterations. This completes the proof.\ \ $\Box$

\section{Proof of Theorem~\ref{t:linfnc}}
\label{ss:linfnc}
{\em Proof.} Similar to the proof of Theorem~\ref{t:l2nc}, we  need to show that there exists a linear model
of the form~\eqref{e:model}, where $\v$ satisfies $\|\A^T\v\|_{\infty}\leq \epsilon$,
$\A$ satisfies the RIP with $\delta_{K+1}=\delta$ for any given $\delta$ satisfying \eqref{e:deltanc},
and $\x$ is $K$-sparse and satisfies~\eqref{e:l2nececounte} with
\beq
\label{e:gammalinf}
0<\gamma<\frac{2\epsilon}{1-\sqrt {K+1 }\delta},
\eeq
such that the OMP algorithm fails to recover the support of $\x$ in $K$ iterations.

We use the same $\A$ and $\x$ (see~\eqref{e:A} and~\eqref{e:x}) as in the proof of Theorem~\ref{t:l2nc},
but instead of~\eqref{e:v2}, we define
\beq
\label{e:vlinf}
\v=-\epsilon\D^{-1}\U\1_{K+1} \in \mathbb{R}^{K+1}.
\eeq
Recall that $\1_{K+1}$ is a $(K+1)$-dimensional column vector with all of its entries being 1.
So by~\eqref{e:A}, we obtain,
$$
\A^T\v=-\epsilon \1_{n},
$$
leading to $\|\A^T\v\|_{\infty}= \epsilon$.

Applying some techniques which are similar to that for deriving  \eqref{e:countleftl2} and \eqref{e:countrightl2}, we have
\beqnn
\max_{i\in\Omega} |\langle \y,\A_i\rangle|=\left(1-\frac{1}{\sqrt{K+1}}\delta\right)\gamma-\epsilon.
\eeqnn
and
$$
\max_{j\in\Omega^c} |\langle \y,\A_j\rangle|=\frac{K}{\sqrt{K+1}}\delta\gamma +\epsilon.
$$
Thus, by~\eqref{e:gammalinf},  we further have
$$
\max_{i\in\Omega} |\langle \y,\A_i\rangle|< \max_{j\in\Omega^c} |\langle \y,\A_j\rangle|.
$$
By line 2 of Algorithm~\ref{a:OMP}, OMP chooses an index in $\Omega^c$ in the first iteration.
Therefore, OMP fails to recover $\Omega$ in $K$ iterations. This completes the proof.~~~~~~~~~~~~~~~~~~~~~~~~~~~ $\Box$

\bibliographystyle{IEEEtran}
\bibliography{ref-RIP}

\end{document}